 \documentclass[10pt,preprint]{aastex}  
 \usepackage{natbib}
\def\ang{\AA}
\def\arcsec{\hbox{$^{\prime\prime}$}}

\def\gapprox{\lower.4ex\hbox{$\;\buildrel >\over{\scriptstyle\sim}\;$}}
\def\lapprox{\lower.4ex\hbox{$\;\buildrel <\over{\scriptstyle\sim}\;$}}

\shortauthors{ASCHWANDEN AND SHIMIZU 2012}
\shorttitle{Spatio-Temporal Evolution of Solar Flares. II.}

\begin{document}

\title{         Multi-Wavelength Observations of the Spatio-Temporal 
		Evolution of Solar Flares with AIA/SDO: 
		II. Hydrodynamic Scaling Laws and Thermal Energies}

\author{        Markus J. Aschwanden }

\affil{		Lockheed Martin Advanced Technology Center,
                Org. ADBS, Bldg.252,
                3251 Hanover St.,
                Palo Alto, CA 94304, USA;
                e-mail: aschwanden@lmsal.com }

\and

\author{        Toshifumi Shimizu }

\affil{		Institute of Space and Astronautical Science, 
		Japan Aerospace Exploration Agency, 3-1-1 Yoshinodai, Chuo, 
		Sagamihara, Kanagawa 252-5210, Japan;
		e-mail: shimizu.toshifumi@isas.jaxa.jp } 

\begin{abstract}
In this study we measure physical parameters of the same set of 155 M and X-class 
solar flares observed with AIA/SDO as analyzed in Paper I, by performing a 
{\sl differential emission measure (DEM)} analysis to determine the flare
peak emission measure $EM_p$, peak temperature $T_p$, electron density $n_p$,
and thermal energy $E_{th}$, in addition to the spatial scales $L$, areas $A$,
and volumes $V$ measured in Paper I. The parameter ranges for M and X-class 
flares are: $\log(EM_p)=47.0-50.5$, $T_p=5.0-17.8$ MK, 
$n_p=4 \times 10^9-9 \times 10^{11}$ cm$^{-3}$, and thermal energies of 
$E_{th}=1.6 \times 10^{28}-1.1 \times 10^{32}$ erg.
We find that these parameters obey the Rosner-Tucker-Vaiana (RTV) scaling law 
$T_p^2 \propto n_p L$ and $H \propto T^{7/2} L^{-2}$ during the peak time $t_p$ 
of the flare density $n_p$, when energy balance between the heating rate $H$ 
and the conductive and radiative loss rates is achieved for a short instant, 
and thus enables the applicability of the RTV scaling law. 
The application of the RTV scaling law predicts powerlaw
distributions for all physical parameters, which we demonstrate with numerical
Monte-Carlo simulations as well as with analytical calculations. A consequence
of the RTV law is also that we can retrieve the size distribution of heating
rates, for which we find $N(H) \propto H^{-1.8}$, which is consistent with
the magnetic flux distribution $N(\Phi) \propto \Phi^{-1.85}$ observed by 
Parnell et al.~(2009) and the heating flux scaling law 
$F_H \propto H L \propto B/L$ of Schrijver et al.~(2004). 
The fractal-diffusive self-organized criticality model in conjunction with
the RTV scaling law reproduces the observed powerlaw distributions and their 
slopes for all geometrical and physical parameters and can be used to predict 
the size distributions for other flare datasets, instruments, and detection 
algorithms.
\end{abstract}

\keywords{Sun: Solar Flares --- Statistics --- Magnetic fields}

\section{INTRODUCTION}

Nonlinear energy dissipation processes governed by self-organized criticality (SOC)
exhibit the ubiquitous powerlaw distribution functions. One of the most 
intriguing questions in this context is still: Why does nature produce powerlaws?
And the very next question is: Can we understand or predict the value of the
powerlaw slopes? While Per Bak attempted to explain the powerlaw nature of
size distributions of SOC avalanches with the functional form of power spectra,
such as the 1/f-noise characteristics that naturally occurs in many systems 
(Bak et al.~1987), we proposed an even more fundamental explanation for the
existence of powerlaws, namely the mere statistical probability of avalanche
sizes that can occur in a SOC system with scale-free properties, which scales with 
$N(L) \propto L^{-d}$, where $d$ is the Euclidean dimension of the SOC system  
(Aschwanden 2012a,b; 2013a,b). This fundamental relationship, which we call the
{\sl scale-free probability conjecture}, predicts directly the size distributions
for avalanche areas $N(A) \propto A^{-2}$ and volumes $N(V) \propto V^{-5/3}$.
This prediction follows from the elementary geometrical relationships that
the area scales with the square of the length scale, $A \propto L^2$, and the
volume scales with the third power of the length scale, $V \propto L^3$.
While the geometric parameters $L$, $A$, and $V$ are space-filling within the
Euclidean dimension $d$, the internal structure of avalanches is highly
inhomogeneous and can be characterized with a fractal dimension $D_d \le d$,
so that the instantaneous avalanche area and volume obey the scaling laws
$A \propto L^{D_2}$ and $V \propto L^{D_3}$, where $D_2$ and $D_3$ are the 
Haussdorf dimensions in 2D and 3D space (Aschwanden 2012a). 
These elementary geometric relationships are probably the simplest {\sl scaling
laws} known in nature. A convenient property of
multiplicative or power-exponent scaling laws is that the powerlaw function of 
a size distribution of one parameter transforms into another, which is the
ultimate reason why we find so many powerlaw distributions in nature.

If we want to understand the powerlaw-like size distributions found for many
observables and physical parameters in solar flares, as well as to understand
the mutual relationships between the powerlaw slopes of these parameters,
we have obviously to look into scaling laws, which is the subject of this study.
In Paper I (Aschwanden, Zhang, and Liu 2013a) we investigated the spatial
and temporal scales of solar flares and found them to be consistent with
the scale-free probability conjecture $N(L) \propto L^{-3}$ for the 3D
Euclidean space, and with a random walk transport process 
$L \propto \tau^{\beta/2}$ with sub-diffusive characteristics ($\beta < 1$).
Analyzing EUV and soft X-ray observations of solar flares involves the
physical parameters of heated plasma that radiates during a flare, which
can be described by the macroscopic parameters of electron temperatures $T_e$,
electron densities $n_e$, ideal gas pressures $p = 2 n_e k_B T_e$, emission
measures $EM$, and thermal energies $E_{th}$. Statistics of such flare
parameters has been gathered for limited samples in the past, such as for
flares observed in soft X-rays and EUV (Pallavicini et al.~1977; Shimizu 1995; 
Feldman et al.~1995a,b, 1996; Porter and Klimchuk 1995; 
Kano and Tsuneta 1995, 1996; 
Metcalf and Fisher 1996; Sterling et al.~1997; Kankelborg et al.~1997; 
Reale et al.~1997; Garcia 1998), or for nanoflares observed in EUV 
(Berghmans et al.~1998; Krucker and Benz 2000; Parnell and Jupp 2000;
Aschwanden et al.~2000;
Aschwanden and Parnell 2002). A universal scaling law involving a hydrostatic
and a magnetic relationship was proposed by Shibata and Yokoyama (1999).
Empirical two-parameter scaling relationships were explored by
Aschwanden (1999). Most of these studies were motivated by testing
the loop heating scaling law of Rosner-Tucker-Vaiana (RTV) 
(Rosner et al.~1978) or by testing whether the energy distribution of
nanoflares is identical with that of larger flares. Several of these studies
suffered from insufficient broadband temperature coverage, which leads to
significant underestimates of the flare energy and even affects the powerlaw
slope of their distributions (e.g., Benz and Krucker 2002; Aschwanden and
Parnell 2002), revealing incompatible results and triggering disputes about 
the true flare energy distribution. A unified flare energy distribution 
was attempted by synthesizing statistics from different observers, 
instruments, and detection algorithms on the same scale 
(e.g., Fig.~10 in Aschwanden et al.~2000), but doubts remained about
the compatibility of different analysis methods and the role of different 
activity levels of the solar cycle. No study has been tackled yet that provides 
a consistent statistics of solar flares from the largest to the smallest flare 
events. The obvious next step in establishing reliable size distributions
of flare energies therefore calls for a single instrument that has broadband
temperature coverage and sufficient cadence to sample the peak time of 
the maximum energy release in flares with adequate time resolution. The answer 
to this call is the AIA/SDO instrument, which allows us to measure all spatial,
temporal, and physical parameters with unprecedented quality. In this Paper II
we perform a {\sl differential emission measure (DEM)} analysis of the same
155 large flare events (M and X GOES class) for which we analyzed the
spatio-temporal parameters in Paper I. Future studies will continue to
smaller classes of flares, and ultimately reveal the true size distribution
of flares from the largest events down to the nanoflare regime.

The content of this paper includes a description of the observations,
the data analysis, and the results in Section 2, theoretical modeling
of the size distributions and correlations in terms of the RTV scaling law 
in Section 3, a discussion of the results in a larger context in Section 4,
and conclusions in Section 5.

\section{OBSERVATIONS, DATA ANALYSIS, AND RESULTS}

\subsection{AIA Observations}

The dataset we are analyzing here is identical with that of Paper I
(Aschwanden, Zhang, and Liu 2013a) and an earlier single-wavelength
study on the spatio-temporal flare evolution (Aschwanden 2012b). 
This dataset consists of 155 solar 
flares that includes all M- and X-class flares detected with 
the {\sl Atmospheric Imaging Assembly (AIA)} onboard the
{\sl Solar Dynamics Observatory (SDO)} 
during the first two years of the mission (from 2010 May 13 to
2012 March 31). All AIA images have a cadence of $\Delta t=12$ s and a
pixel size of $\Delta x=0.6\arcsec \approx 435$ km, which corresponds
to a spatial resolution of $2.5\Delta x = 1.5\arcsec \approx 1100$ km.
We analyzed about 100 time frames for each flare in seven coronal 
wavelength filters (94, 131, 171, 193, 211, 304, 335 \ang ) of AIA/SDO 
(Lemen et al.~2012; Boerner et al.~2012), which amounts to a total 
number of $\approx 10^5$ images or $\approx 2$ Terabytes of AIA data.

In Paper I we measured the following parameters for each flare:
The (time-integrated) flare area $A$ at the peak time $t_p$ of the GOES 
soft X-ray flux (which approximately corresponds to the density peak time
and to the end time of nonthermal hard X-rays according to the Neupert 
effect); the length scale 
$L=\sqrt{A/\pi}$ defined by the radius of an equivalent circular flare
area $A$; the flare volume $V=(3\pi/2) L^3$ (defined by a hemisphere
with radius $L$), the area fractal dimension $D_2=\log{(a)}/\log{(L)}$ (defined
by the instantaneous fractal flare area $a$ at time $t_p$), the flare 
duration $\tau$ (defined by the soft X-ray rise time $\tau=t_p-t_{start}$,
which roughly corresponds to the duration of nonthermal hard X-ray
emission according to the Neupert effect), the diffusion coefficient 
$\kappa$ and the diffusion or spreading exponent $\beta$ (defined by
the generalized diffusion equation $L = \kappa \tau^{\beta/2}$), 
and the maximum expansion velocity $v_{max}$, all measured for each
of the seven wavelengths. In this study we measure in addition the
peak emission measure $EM_p$ and the peak temperature $T_p$, which
characterize the peak of the differential emission measure (DEM)
distribution function at the peak time $t_p$ of soft X-ray emission, 
obtained from the six coronal AIA wavelength fluxes $F_{\lambda}$ 
at the flare peak time $t_p$ (without the 304 \ang channel).

\subsection{AIA Temperature Response Functions}

AIA has six wavelength filters that are sensitive to highly ionized
iron lines at coronal temperatures (94, 131, 171, 193, 211, 335 \ang )
and one (304 \ang ) that is sensitive to chromospheric temperatures.
The contribution of spectral lines and continuum emission to the EUV
channels of AIA/SDO are listed in O'Dwyer et al.~(2010), and MHD
simulations are provided in Martinez-Sykora et al.~(2011). 
The response functions $R(T)$ of these seven filters are shown elsewhere
(e.g., Fig.~13 in Lemen et al.~2012) and can be obtained with the 
{\sl Solar SoftWare (SSW)} procedure {\sl AIA$\_$GET$\_$RESPONSE}.
The temperatures range extends to as low as $T \approx 0.1$ MK for 
the 304 \ang\ filter, and covers the range of $T_e \approx 1-20$ MK 
for the coronal filters (131, 171, 193, 211, 335, and 94 \ang ).
Most filters have a dual response to two temperature ranges, 
which makes the inversion of the differential emission measure (DEM)
distribution more ambiguous. The AIA instrument is described in
Lemen et al.~(2012) and the calibration is detailed in Boerner
et al.~(2012). A major update in the AIA instrumental response functions
since launch occurred on February 13, 2012, which includes updated
emissivities according to the CHIANTI (Version 7) model, and an
empirical correction of the 94 and 131 \ang\ sensitivities in the
lower temperature range ($log(T) \lapprox 6.3$) due to missing 
Fe XVIII, XI, and X lines in the CHIANTI model, as reported earlier
(Aschwanden and Boerner 2011; Fig.~10 therein). In the present study 
we used the updated standard response functions that are available in the
{\sl Interactive Data Language (IDL)} based {\sl Solar Software (SSW)}
(status of December 2012).

\subsection{Differential Emission Measure Analysis}

We determine the {\sl differential emission measure (DEM)} distribution
function for each of the $\approx 100$ time intervals (per flare) for
the 155 selected M- and X-class flares, in order to obtain the
wavelength-independent peak emission measure $EM_p$ and DEM peak temperature
$T_p$ per event. An example of a dataset for one event (observed on
2011 March 23, 2:00-2:30 UT, a GOES M1.4 class flare) is shown in Fig.~1.
The GOES time profiles for the soft (1-8 \ang ) and hard (0.5-4 \ang )
channels are shown in the top panel of Fig.~1, with the GOES start
and end times (vertical dashed lines in Fig.~1) and flare peak time
$t_p$ (vertical solid line in Fig.~1) indicated. The accompanying
AIA/SDO time profiles for all seven filters are shown in the second
panel of Fig.~1. From the colored time profiles one can identify that
the 171, 211, and 193 \ang\ channels peak before the GOES soft channel,
while the 131, 94, 335, and 304 channels peak after the GOES channel.
Thus, the EUV channels peak over a range of time (02:12-02:24 UT).
With the DEM analysis we deduce the time evolution of the total emission
measure $EM(t)$ (integrated over the flare area) and electron temperature 
$T_e(t)$.   

Our observational constraints are the six EUV fluxes $F_{\lambda}(t)$, 
since we ignore the chromospheric channel $\lambda=304$ \ang\ in DEM fits. 
In the following, the flux $F_\lambda$ (DN s$^{-1}$) always refers 
to the total flux integrated over the flare area, 
and the total emission measure $EM$ (cm$^{-3}$) is integrated over 
the entire flare volume. 
We subtract a preflare-background flux $F_{\lambda}(t_b)$ in each channel,
which is believed to originate from flare-unrelated emission in the
active region, determined from the minimum flux in the time interval
$t_{start} < t_b < t_{peak}$ between the GOES flare start time 
$t_{start}$ and the GOES flare peak time $t_p$. 
These background-subtracted loop fluxes in each of the six coronal
wavelength filters can be related to the DEM distribution 
function $dEM(T)/dT$ of the flare at time $t$ by
\begin{equation}
        F_{\lambda}(t)-F_{\lambda}(t_b) 
	= \int {dEM(T,t) \over dT} R_{\lambda}(T) \ dT
                     = \sum_k EM(T_k,t) R_{\lambda}(T_k) \Delta T_k \ ,
\end{equation}
where $R_{\lambda}(T)$ is the instrumental temperature response function 
of each filter $\lambda$. The particular functional shape of the 
DEM function $dEM(T,t)/dT$ is unknown and can be quite complex, based
on DEM reconstructions of solar flare multi-wavelength observations
(e.g., Battaglia and Kontar 2012; Graham et al.~2013). 
Also, the DEM reconstruction from AIA data is not necessarily reliable,
based on simulations with known DEMs from MHD simulations,
although acceptable $\chi^2$-values may be obtained (e.g., Testa et al.~2012). 
Nevertheless, for sake of simplicity, the temperature peak of a DEM 
distribution can be characterized with a minimum of three free parameters, 
such as with a Gaussian function in the logarithm of the temperature,
\begin{equation}
        EM(T, t_i) = EM_i \exp{\left(- {[\log(T)-\log(T_i)]^2
                \over 2 \sigma_{i}^2}\right)} \ ,
\end{equation}
which has three free parameters for every time $t_i$. At the density peak time
$t_i=t_p$ of the flare, this defines the peak emission measure $EM_p$,
the DEM peak temperature $T_p$, and the Gaussian temperature 
width $\sigma_p$. The observed functions $F_{\lambda}(t)$ can
then be fitted simply by calculating the convolution of the
DEM function $dEM(T)/dT$ with the filter response functions 
$R_{\lambda}(T)$ (Eq.~1) for a set of discretized temperatures
$T_i, i=0,...,n_T$ and Gaussian widths $\sigma_j, j=1,...,n_\sigma$,
while the emission measure $EM_p$ is just a constant that can be
obtained from the ratio of the left-hand and right-hand side 
terms of Eq.~1, using the median value among the six wavelengths. 
The choice of a Gaussian is one of the simplest functions to model a
DEM, and thus most robust, but we caution that it may not be
adadequate to characterize some broadband DEMs. 

The best fit of the parameters $T_p$ and $\sigma_p$ is 
normally obtained by a chi-square fit, which requires knowledge of
the expectation value of the uncertainty $\sigma_F$ in the count 
rate $F$ in each filter. For the Poisson statistics of the photons,
which is for AIA approximately equal to the the counts 
$C = F \times t_{exp}$ recorded during the exposure time $t_{exp}$, 
the expectation value of the uncertainty is 
$\sigma_C = \sqrt{C}$ (e.g., see Aschwanden and Boerner 2011 
for a DEM analysis of coronal loops using AIA data).
In the case of total emission measure modeling of flares as applied
here, photon statistics can be neglected, because the largest 
uncertainty comes from the inadequacy of the chosen functional form 
of the DEM, and from the estimation of the subtracted preflare
background, which is difficult (or impossible) to quanitfy {\sl a priori}, 
since the background has large spatial and temporal variations that cannot
easily be separated from flare-related EUV emission. However, we
can straightforwardly quantify a measure of the goodness-of-fit by
the ratios of the fitted to the observed fluxes, in terms of a
mean and r.m.s. standard deviation $\sigma_{dev}$,
\begin{equation}
	\sigma_{dev} = \left[ {1 \over n_{\lambda}} 
			\sum_\lambda (f_{fit,\lambda}-f_{obs,\lambda})^2
			\right]^{1/2} \ .
\end{equation}
Minimizing this fitting criterion in each DEM fit (for each time $t$) 
we obtain the time evolution of the peak emission measure $EM_p(t)$,
peak temperature $T_p(t)$, and Gaussian temperature width $\sigma_p(t)$.
We found a most robust optimization by using the goodness-of-fit
measure $\sigma_{dev}$ as a weighting factor $w_{ij}=1/\sigma_{dev,ij}^2$ 
in averaging all trial values of $T_i$ and $\sigma_j$,
\begin{equation}
	T_p = {\Sigma_i^{n_T} \Sigma_j^{n_\sigma} T_i w_{ij} \over
	       \Sigma_i^{n_T} \Sigma_j^{n_\sigma} w_{ij} } \ ,
\end{equation}
\begin{equation}
	\sigma_p = {\Sigma_i^{n_T} \Sigma_j^{n_\sigma} \sigma_j w_{ij} \over
	       \Sigma_i^{n_T} \Sigma_j^{n_\sigma} w_{ij} } \ .
\end{equation}
The time evolution of the emission measure $EM_p(t)$ and electron
temperature $T_p(t)$ at the peak of the DEM are shown in the third
panel of Fig.~1. The logarithmic (Gaussian) temperature width $\sigma_p$  
of the DEM at the peak time is found to have a mean and standard 
deviation of $\sigma_p = 0.50\pm0.13$ (for the entire set of 155 flares), 
but is not correlated with
either the temperature $T_p$ or emission measure $EM_p$. So, the
average full width at half maximum (FWHM) of the flare DEMs is 
$log({\rm FWHM}_T)\approx 2.35 \sigma_p \approx 1.18$. This is quite a
broad DEM function. For instance, for a typical flare peak temperature of
$T_p \approx 10$ MK we have a FWHM temperature range from 
$T_1=T_p-{\rm FWHM}_T/2=2.5$ MK to $T_2=T_p+{\rm FWHM}_T/2=40$ MK.

Using furthermore the spatial length scale $L(t)$ and volume $V(t)$
as determined in Paper I, we can then also derive the time evolution 
of the average electron density (averaged over the flare volume),
\begin{equation}
	n_e(t) = \sqrt{ EM_p(t) \over V(t)} \ ,
\end{equation}
where the flare volume $V(t)$ is approximated by a hemispheric
geometry, $V(t)=(2/3)\pi L(t)^3$. In addition we infer also the
time evolution of the thermal energy,
\begin{equation}
	E_{th}(t) = 3 \ n_e(t) k_B T_p(t) V(t) \ .
\end{equation}
These time evolutions are also shown in the third panel of Fig.~1,
where it can be clearly discerned that all quantities, i.e., the
emission measure, the temperature, the density, and the thermal
energy monotonically increase before the flare peak, while all
drop after the flare peak, except for the flare size $L(t)$ and
volume $V(t)$. The fitting quality of this example is quite satisfactory,
as it can be seen from the ratio of the fitted to the observed
fluxes, with a mean and standard deviation of
$f_{fit}/f_{obs}=1.05 \pm 0.11$ (listed in Fig.~1, bottom right). 
In other words, the DEM at the
flare peak time can be characterized by a Gaussian function that
reproduces all 6 coronal EUV fluxes with an accuracy of $\approx 15\%$
(except for the 304 \ang\ channel, which is of chromospheric origin
and ignored here).
An evolutionary temperature-density phase diagram $T_p(n_p)$ is shown
in the lower left panel in Fig.~1, revealing a density and
temperature increase that closely follows the RTV law (dashed
diagonal line) during the flare rise time, while it deviates from
the RTV equilibrium during the beginning of the cooling phase
as expected from hydrodynamic simulations (e.g., Jakimiec et al.~1992; 
Sterling et al.~1997; Aschwanden and Tsiklauri 2009). 
 
In the remainder of the paper we will ignore the time evolution
and deal with statistics of the physical parameters measured at the 
GOES peak time $t_p$ only, which approximately equals the 
peak time of the flare emission measure $EM_p$ or density $n_p$. 
We label the parameters briefly as $T_p = T_e(t=t_p)$ (not to be
confused with the temperature maximum time $t_m$ that occurs earlier
than the density peak time $t_p$), 
$EM_p = EM(t=t_p, T_e=T_p)$, $n_p = n_e(t=t_p, T_e=T_p)$, and
$E_{th} = E_{th}(t=t_p, T_e=T_p)$. In the example shown in Fig.~1,
these values amount to $EM_p=10^{49.42}$ cm$^{-3}$,
$T_p=14.13$ MK, $log(\sigma_T)=0.49$, $L_p=12.7$ Mm,
$n_p=1.17 \times 10^{11}$ cm$^{-3}$, and $E_{th}=2.92 \times
10^{30}$ erg. 

The best-fit parameters of the DEM analysis ($T_p, \sigma_p, EM_p,
E_{th}$, and $F_{fit}/F_{obs}$) are tabulated in Table 2 for all
155 analyzed events. Note that the length scale $L_p$ listed in Table 2
refers to the radius of the time-integrated flare area above some
flux threshold, which may deviate somewhat from the values $L$ 
measured from the radial expansion with preflare-area subtraction 
in Aschwanden (2012b). We identify a few events that represent outliers 
regarding the peak temperature $T_p \lapprox 5$ MK 
(events \#11, 18, 90, 100), emission measure $\log(EM_p) \lapprox 47.5$
(events \#7, 26, 90, 139), or DEM fit quality ($F_{fit}/F_{obs} \gapprox 1$)
(event \#90), which partially are affected by preflare background problems
or other data irregularities.  

\subsection{Observed Size Distributions}

The size distributions of the physical parameters of the 155 analyzed
flares are shown in Fig.~2, which are all wavelength-independent,
because the DEM analysis synthesizes the contributions from all 
different wavelengths into an instrument-independent DEM distribution
function. Besideds the powerlaw part on the right side of the distribution,
there is also a rollover at the left side due to incomplete sampling.

The geometric parameters
show powerlaw distributions with the universal values for their
slopes, i.e., $\alpha_L=2.80\pm0.16$ (Fig.~2a, predicted as $\alpha_L=3.00$ 
from the FD-SOC model) for length scales within a range of $L=4-64$ Mm, 
in agreement with the averaged values from all wavelengths and flux 
thresholds in Paper I, namely $\alpha_L=3.20\pm0.71$. 
The powerlaw slope of the flare volumes is found to be 
$\alpha_V=1.62\pm0.04$ (Fig.~2b, predicted as $\alpha_V=5/3 \approx 1.67$ 
from the FD-SOC model) for volumes in a range of $V=2 \times 10^{26}-5 \times 
10^{29}$ cm$^{-3}$. 

For the physical parameters we find from the DEM analysis the following
ranges and powerlaw slopes: 
emission measure $EM_p=10^{47.0}-10^{50.5}$ cm$^{-3}$, 
$\alpha_{EM}=1.78\pm0.03$ (Fig.~2c);
electron temperature $T_P \approx 5.0-17.8$ MK (Fig.~2d);
electron density $n_P \approx 4 \times 10^9 - 9 \times 10^{11}$ cm$^{-3}$, 
$\alpha_{n}=2.15\pm0.17$ (Fig.~2e); and
thermal energy $E_{th}=1.6 \times 10^{28}-1.1 \times 10^{32}$ erg,
$\alpha_{Eth}=1.66\pm0.13$ (Fig.~2f).
We note that the powerlaw slopes of the flare volume 
($\alpha_V=1.62\pm0.04$. Fig.~2b), peak emission measure  
($\alpha_{EM}=1.78\pm0.03$, Fig.~2c), and thermal energy 
($\alpha_{Eth}=1.66\pm0.13$, Fig.~2f) are almost identical, and thus these 
three parameters are almost proportional to each other. 
By defining a powerlaw scaling between two parameters $x$ and $y$, i.e.,
$x \propto y^\gamma$ with $N(x) \propto x^{-\alpha_x}$,
$N(y) \propto y^{-\alpha_y}$, and $\gamma=(\alpha_x-1)/(\alpha_y-1)$,
we find the following powerlaw 
relationships between the three parameters $V_p$, $EM_p$, and $E_{th}$, 
\begin{equation}
	EM_p \propto V^{\gamma_{EM}} \ , \qquad
	\gamma_{EM} = {\alpha_V - 1 \over \alpha_{EM} - 1} = 0.80 \pm 0.07 \ ,
\end{equation}
\begin{equation}
	E_{th} \propto V^{\gamma_{Eth}} \ , \qquad
	\gamma_{Eth} = {\alpha_V - 1 \over \alpha_{Eth} - 1} = 0.85 \pm 0.07 \ ,
\end{equation}
\begin{equation}
	E_{th} \propto EM_p^{\gamma_{E}} \ , \qquad
	\gamma_{E} = {\alpha_{EM} - 1 \over \alpha_{Eth} - 1} = 1.23 \pm 0.29 \ .
\end{equation}
These relationships allow us to predict the thermal flare energy $E_{th}$ from
the observed peak emission measure $EM_p$ directly, without need of spatial 
observations, which may be used in the case of stellar observations.
Alternatively, we can predict the geometric volume $V_p$ (or the spatial
size of the flare) from the observed emission measure $EM_p$, which may be used 
for both solar and stellar non-imaging observations, supposed that flares on 
the Sun and on stars obey the same scaling law.

The size distributions of peak temperatures and densities, 
$N(T_p)$ and $N(n_p)$, cannot straightforwardly be predicted from 
two-parameter correlations, since we expect three-parameter relationships 
$(T_p, n_p, L_p)$ for RTV-type scaling laws (see Section 3).

The size distributions of peak fluxes $F_{\lambda}$ for the AIA channels
with wavelengths $\lambda=$ 94, 131, 171, 193, 211, 304, and 335 \ang\
have an approximate powerlaw slope of $\alpha_F \approx 2.0$ (Table 1). 
A breakdown of the
fitted powerlaw slopes $\alpha_{\lambda}$ by wavelengths $\lambda$ and 
flux thresholds for flare area detection $q_{thresh}=$ 0.01, 0.02, 0.05,
0.1, and 0.2 is listed in Table 1, which yield a mean and standard
deviation of $\alpha_{\lambda}=2.06\pm0.13$. The lack of any significant
wavelength dependence implies a very good correlation between fluxes
in different wavelengths, which is expected for broadband DEM functions
and/or broadband instrumental response functions.

\subsection{Observed Flux-Flux Correlations}

Optically-thin emission in soft X-ray and EUV lines are expected to
correlate with the flare volume, since the observed fluxes $F_{\lambda}$
scale with the emission measure $EM$ (Eq.~1), and thus with the total
flare volume $V$, since $EM = \int n_e^2 dV$ (Eq.~6). We show the
flux-volume correlations in form of scatterplots and linear regression
fits in Fig.~3 (top and second row). 
The linear regression fits use the {\sl orthogonal reduced major axis} 
method (Isobe et al.~1990). The Pearson cross-correlation coefficient
of the AIA flare peak fluxes $F_p$ with the inferred flare volumes $V_p$
(Paper I) vary from the lowest value $ccc=0.18$ for the 193 \ang\  
wavelength to $ccc=0.70$ for the 335 \ang\ wavelength. The relatively
low values for the 171 and 193 \ang\ channels probably result from the
effects of EUV dimming, which cause an underestimate of cool ($T_e=1-2$ MK) 
plasma due to coronal mass ejections (Aschwanden et al.~2013a;
Paper I, Section 4.4). Interestingly, the linear regression coefficient
$\gamma$ obtained from the correlation $F_{AIA} \propto V^\gamma$
behaves similar to the cross-correlation coefficient,
namely the slope is highest and nearest to proportionality for
335 \ang\ ($\gamma_{335}=0.59\pm0.05$ and $ccc=0.70$), while it is
lowest for 193 \ang  ($\gamma_{193}=0.16\pm0.08$ and $ccc=0.16$).
Thus, both the cross-correlation coefficient $ccc$ and the linear
regression slope $\gamma_{\lambda}$ are a measure of the flux-volume
proportionality. We will discuss the wavelength-dependent flux-volume
relationships in Section 3.5. 

The GOES flux (of the soft 1-4 \ang\ channel) is often used to define
the flare magnitude (such as in the selection of M- and X-class flares used
here). We show the correlations between the EUV peak fluxes $F_{AIA,\lambda}$
and the GOES peak fluxes $F_{GOES}$ measured at the flare peak time $t_p$,
both preflare-background subtracted, in Fig.~3 (third and bottom
row). We see that the best correlations with the GOES soft X-ray flux occur 
for those EUV channels that are sensitive to high temperatures, 
such as 193 \ang\ (ccc=0.82), 131 \ang\ (ccc=0.79), and 94 \ang (ccc=0.74),
while it is lowest for chromospheric temperatures as seen with 304 \ang\
(ccc=0.48). The linear regression fits show correlations between
the AIA and GOES fluxes, $F_{AIA, \lambda} \propto 
F_{GOES}^{\delta_{\lambda}}$, with $\delta_\lambda=1.04\pm0.13$, 
excluding the 304 \ang\ channel. The three hottest EUV channels 
(193, 131, 94 \ang) thus represent some proxies of the GOES fluxes
or flare magnitude, as expected to some degree from the GOES 
high-temperature response function (White et al.~2004). 
A recent study used the 193 \ang\ channel on the
STEREO spacecraft, which besides the peak sensitivity to $T_e \approx 1.5$
MK plasma includes also an Fe XXIV line at 192 \ang\ that enables a secondary
sensitivity at 15 MK, to estimate the GOES magnitude of occulted or
behind-the-limb flares (Nitta et al.~2013). 

\section{THEORETICAL MODELING}

In this study we determined physical parameters of flares, such as
electron temperatures $T_e$ and electron densities $n_e$, using
a multi-wavelength {\sl differential emission measure (DEM)} analysis
that provides the DEM peak emission measure $EM_p$ and peak temperature $T_p$
at the flare density peak time $t_p$. In order to understand the underlying
physical scaling laws, their statistical distributions, and correlations,
we have to relate these physical parameters to the geometric parameters 
(length scale $L$, flare area $A$, flare
volume $V$, and fractal dimension $D_d$) determined in Paper I.
Our theoretical model is based on the RTV (Rosner, Tucker, and Vaiana
1978) scaling law (Section 3.1), which provides a useful tool that can 
be applied at the flare peak time to multi-loop flare geometries (Section 3.2)
and can predict parameter correlations and powerlaw distributions of
the observed parameters (Section 3.3-3.5). 

\subsection{The RTV Scaling Law}

The RTV scaling law (Rosner et al.~1978) describes a hydrostatic
equilibrium solution of a coronal loop that is steadily and spatially 
uniformly heated, has a constant pressure, and is in equilibrium between 
the volumetric heating rate and the losses by radiation and thermal
conduction, yielding a scaling law between the loop maximum temperature 
at the apex, the (constant) pressure, and the loop half length,
as well as a scaling law for the (constant) volumetric heating rate. 
Although this scaling law is generally applied to one-dimensional
(1-D) coronal loops, the assumption of steady heating and spatial
uniformity of the heating function is often questioned, because
observations and hydrodynamic modeling suggest impulsive heating 
functions (e.g., Warren et al.~2003) and non-uniform (footpoint) 
heating (e.g., Aschwanden et al.~2001). 

Nevertheless, despite the observed violation of the steady-state 
assumption, the application of the RTV scaling law is probably most
adequate at one particular time of an impulsively-heated 
coronal loop, when it reaches the density peak 
$n_p=n_e(t=t_p)$. At this particular time 
the energy balance is approximately fulfilled, where the
heating rate just matches the energy losses due to thermal conduction
and radiative losses. Before the time $t_p$, heating dominates
over the losses and the temperature rises, while after this time at
$t \ge t_p$ the energy losses dominate over heating and the loop
temperature drops. 
We illustrate this hydrodynamic behavior in Fig.~4, where
the temperature and density evolution is depicted according to a
hydrodynamic simulation with an impulsive heating function (see
details in Aschwanden and Tsiklauri 2009).  In the evolutionary
temperature-density phase diagram (Fig.~4, right panel),
a phase diagram that has been pioneered extensively with earlier 
flare data (e.g., Jakimiec et al.~1992; Sylwester et al.~1990, 1993; 
Sylwester 1996), we can overlay the
RTV equilibrium solution $T_e \propto n_e^{1/2}$ for the simulated
loop with a half length of $L_{loop}=5.5 \times 10^9$ cm.
If the loop would be very slowly heated, near-equilibrium energy 
balance could be achieved and the temperature and density rise 
should follow the RTV curve. However, the evolutionary curve of the
hydrodynamic simulation shows that the loop temperature is always
higher than the RTV equilibrium value during the heating-dominated
phase, while it is lower during the cooling-dominated phase
(Fig.~4, right panel). The RTV solution predicts the correct loop
temperature and density only at one point in time, near the peak 
density time $t=t_p$, where the peak density $n_e=n_p$ is reached 
and the temperature dropped to about half of the maximum temperature,
$T_p \approx T_m/2$. Thus the RTV law is a useful predictor for the
density $n_p$ and temperature $T_p$ when the emission of the loop
is brightest, since the emission measure $EM_p$ scales with the squared
density $n_e$ and line-of-sight column depth $L_z$, i.e., 
$EM_p = n_p^2 L_z$.

While the foregoing argument was made for a single loop in an active region,
we are using now the same argument of the applicability of the RTV scaling
law for the flare peak time $t_p$, when the thermal emission of a
flare in soft X-ray wavelengths is brightest, such as during the
GOES peak time $t_p$ of the flare. Comparing the evolutionary
phase diagram $T_e(n_e)$ of the total flare emission of an observed
flare (Fig.~1, bottom left panel) with the hydrodynamic simulation of
an impulsively-heated single loop (Fig.~4, right panel), we see a
fairly similar evolution, although the flare may consist of multiple 
loops.  

\subsection{Multi-Loop Flare Geometries}

Can we apply the RTV scaling law to multi-loop geometries?
Spatial high-resolution observations from TRACE and AIA/SDO clearly
demonstrate the multi-loop structure of solar flares. One particular
flare, the Bastille-Day-2000 flare has been modeled in detail and
a multitude of at least $\gapprox 100$ individual postflare loops
have been identified for this X5.7 GOES-class flare (Aschwanden and
Alexander 2001). Even the simplest and smallest nanoflares appear
to be composed of multiple loops (e.g., Aschwanden et al.~2000).
Such multi-loop geometries can be modeled most simply by arcades
of loops, straddling along a neutral line, as visualized in Fig.~5.
In Paper I, however, we measured the projected flare area
$A(t)$ and defined a length scale 
\begin{equation}
	L = \sqrt{A/\pi} \ ,
\end{equation}
that corresponds
to the radius of an equivalent circular area $A$. The question arises
now how can we relate this flare length scale $L$ to the loop half
length $L_{loop}$ used in the RTV scaling law, and to multi-loop 
models of flares.

The semi-cylindrical multi-loop arcade model shown in Fig.~5
has a projected area that can be characterized with a rectangular 
shape with a length $l$ and width $w$, yielding an area of
$A=(l w)$, or a radius $L = \sqrt{A/\pi} = \sqrt{l w / \pi}$ 
of an equivalent circular area $A$. The loop half
length $L_{loop}$ for the largest loops contained in the arcade
scale as $L_{loop}=(\pi/2)(l/2)$ for single-loop cases, to 
$L_{loop}=(\pi/2)(w/2)$ for large multi-loop arcades, so the
geometric mean is a good approximation,
\begin{equation}
	L_{loop} = {\pi \over 2} L = {\sqrt{\pi l w} \over 2} \ .
\end{equation}

In principle, a multi-loop arcade consisting of an array of loops with
different (half) lengths $L_{loop}$ as a function of the radial distance 
$r$ from the neutral line could be modeled by a superposition of
RTV loops, according to the scaling laws for the heating rates $H$
and apex temperatures, which can then be summed up to
produce a DEM distribution for each time of a flare, and can then be
related to the overall length scale $L$ of the flare. However, since
the spatial distribution of the heating rate $H$ is unknown and is likely 
to be spatially non-uniform, no simple model can be made with certainty 
{\sl a priori}. Our working assumption is that loops with a half length of
$L_{loop}=(\pi/2)L$ demarcate the brightest loops in the flare arcade,
from which the flare area $A=\pi L^2$ is measured, and thus this
loop (half) length $L_{loop}=(\pi/2)L$ is the most relevant parameter
in the application of the RTV law to estimate the peak emission
measure $EM_p$ and peak temperature $T_p$ for a flare with a
hemispheric volume $V=(2 \pi/3) L^3$ and radius $L$.  

\subsection{Observational Test of the RTV Scaling Law}

The RTV law (Rosner et al.~1978) specifies a 3-parameter relationship between
the flare peak temperature $T_p$, the (spatially averaged) peak electron 
density $n_p$, and length scale $L_p=L_{loop}/(\pi/2)$, measured
at the flare peak time $t_p$. Thus, we can express the RTV law as a function 
of these three parameters $(T_p, n_p, L_p)$, by inserting $L_{loop}=
(\pi/2)L_p$ (Eq.~12), and defining $n_p=n_{apex}$ 
and $T_p=t_{apex}$, to predict each of the three parameters
as a function of the other two,
\begin{equation}
	T_p = c_1 \ n_p^{1/2} \ L_p^{1/2} \ , \qquad
	c_1 = 1.1 \times 10^{-3} \ ,
\end{equation}
\begin{equation}
	n_p = c_2 \ T_p^{2} L_p^{-1} \ , \qquad 
	c_2 = 8.4 \times 10^5 ,
\end{equation}
\begin{equation}
	L_p = c_3 \ T_p^{2} n_p^{-1} \ , \qquad
	c_3 = 8.4 \times 10^5 \ .
\end{equation}
These are predicted 3-parameter correlations that can be tested
with our data. Another useful parameter is the total
emission measure $EM_p$, defined by the integral over the
flare volume $V$,
\begin{equation}
	EM_p = \int n_p^2 dV = n_p^2 V = n_p^2 ({2 \pi \over 3} L_p^3 )
	     = c_4 \ T_p^4 L_p \ , \qquad
	       c_4 = 1.48 \times 10^{12} .
\end{equation}
Furthermore, we will also determine the distribution of thermal energies,
\begin{equation}
	E_{th}  = 3 n_p k_B T_p V_p
		= c_5 \ T_p^3 L_p^2 \ , \qquad
		  c_5 = 7.3 \times 10^{-10} ,
\end{equation} 
which can be expressed as a function of the peak 
temperature $T_p$ and length scale $L_p$ by substituting the 
RTV scaling law for the density $n_p$ (Eq.~14).
In addition we have the RTV heating rate scaling law, 
\begin{equation}
        H \approx c_6 \ T_{max}^{7/2} L_{loop}^{-2} 
		= c_6 \ T_p^{7/2} L_p^{-2} \ , \qquad
                   c_6 = 0.95 \times 10^{-6} 
\end{equation}

Our data provide the measurements of the independent parameters $T_p$,
$n_p$, and $L_p$, which allows us to test the applicability of the RTV
law. In Fig.~6a we show the RTV-predicted flare peak temperature
$T_{RTV}=c_1 \sqrt{n_p L_p}$ (Eq.~13) as a function of the observed flare
peak temperature $T_p$ and find a good agreement within
a factor of $T_{RTV}/T_{obs}=1.05\pm0.38$, which means that the RTV
law predicts the flare temperature with an accuracy of $\approx 40\%$
in the range of $T_p\approx 4-20$ MK.

Similarly we test the predicted peak densities $n_{RTV}$ (Eq.~14)
in Fig.~6b and obtain agreement within a factor of two, i.e.,
$n_{RTV}/n_{obs}=1.3\pm2.1$. The predicted loop lengths $L_{RTV}$
(Eq.~15) are shown in Fig.~6c and agree by the same factor,
$L_{RTV}/L_{obs}=1.3\pm2.1$. The peak emission measures $EM_p$ 
(Eq.~16) are shown in Fig.~6d and agree within $log(EM_{RTV}/EM_{obs})
=0.44\pm0.50$, which corresponds to a factor of $10^{0.44}\approx 2.7$.
The thermal energies $E_{th}$ (Eq.~17) are shown
in Fig.~6e and agree within $E_{RTV}/E_{obs}=1.3\pm2.1$.

We conclude that the RTV law represents a physical model that is
adequate to explain the relationship between geometric flare 
parameters $(L_p, A_p, V_p)$ and the hydrodynamic fluid parameters
of the electron temperature $T_p$ and density $n_p$ at the time of
the flare peak. This corroborates our assumption that energy balance
between heating and cooling processes is achieved during the flare
peak time, even in the case of a dynamic evolution that is different
from the steady-state condition under which the RTV law was derived
originally.

\subsection{Size Distributions Modeled with the RTV Law}

In the next step we like to understand the powerlaw slopes of the 
observed size distributions (Fig.~2) in terms of the RTV scaling law. 
The histogrammed distribution functions show powerlaws at the right-hand
side of the distributions, but reveal also a rollover at the left-hand
side, which is primarily a consequence of incomplete sampling of small 
events, as well as truncation effects that need to be included in 
modeling of the size distributions. In Fig.~7 we show the same
size distributions as in Fig.~2, but juxtapose also scatterplots
of the various parameters (x-axis) with the length scale (y-axis),
which clearly show how the rollover in the size distributions relate to
truncation effects in the scatterplots. There are two truncation effects
that we include in modeling the size distributions: (i) a lower limit 
$H_0 \approx 0.04$ erg cm$^{-3}$ s$^{-1}$ of the heating rate distribution
$N(H)$ (shown as dotted vertical line in Fig.~7d), and (ii) a lower limit
$EM_0=10^{49}$ cm$^{-3}$ of the emission measure threshold that results
from the GOES flux threshold of M1.0 flares (shown as dashed vertical line
in Fig.~7e). The truncation boundaries in the two-parameter scatterplots
of flare peak temperatures (Fig.~7b), peak densities (Fig.~7c), and
thermal energies (Fig.~7f) that result from these two limits $H_0$ and
$EM_0$ are calculated from the RTV law in Appendix A (Eqs.~A1-A4) and 
are indicated in all panels of Fig.~7 as dotted and dashed lines. We see
that these truncation boundaries, caclulated analytically from the RTV
law, demarcate the observed ranges of datapoints in the scatterplots 
quite well. Obviously, they also affect the powerlaw slopes of the
parameter distributions.

There are two methods to corroborate the RTV scaling and the resulting
truncation effects theoretically: (i) by a Monte-Carlo simulation (which
we describe in the following and show in Fig.~8), and (ii) by analytical 
calculations (which we derive in Appendix A).

A Monte-Carlo simulation can easily be performed by generating two sets
of parameters according to the two prescribed distributions of
length scales $L_p$ (following the scale-free probability conjecture, as
verified by measurements in Paper I), 
\begin{equation}
	N(L_p) \propto L_p^{-3} \ ,
\end{equation}
and volumetric heating rates $H$, for which we also assume a powerlaw
distribution for mathematical convenience (for a justification see also 
the analytical derivation of a scaling law for the coronal heating rate 
in Appendix B), 
\begin{equation}
	N(H) \propto H^{-\alpha_H} \ .
\end{equation}
A sample of values $x_i, i=1,...,n$ that obey a powerlaw 
distribution $N(x) \propto x^{-\alpha_x}$ with a lower cuotff of 
$x_0 \le x$ can simply be generated by the relationship (Aschwanden 2012a),
\begin{equation}
	x_i = x_0 (1 - \rho_i)^{1/(1-\alpha_x)} \ ,
\end{equation}
where $\rho_i$ is a random number drawn from the interval 
$\rho=[0,...,1]$ with a uniform probability distribution $N(\rho)
=const$ that can be obtained from a random generator. This way
we simulate a set of values $L_i$ and $H_i$ that form the 
probability distributions $N(L_p) \propto L_p^{-\alpha_L}$
(Eq.~19) and $N(H) \propto H^{-\alpha_H}$ (Eq.~20), were we choose
the lower limits $L_{min}=4 \times 10^8$ cm and $H_0=0.04$ erg cm$^{-3}$
obtained from the observed distributions (Fig.~7), and powerlaw
indices $\alpha_L=3.0$ given by the scale-free probability conjecture 
(Eq.~19) and $\alpha_H=1.8$ that is empirically found to fit the 
observations (Fig.~7d).
From these independent parameters $L_i$ and $H_i$ we obtain then
the temperature values $T_i=(L_i^2 H_i / c_6)^{2/7}$ according
to the RTV energy rate scaling law (Eq.~18), and the density values
$n_i = c_2 T_i^2/L_i$ from the RTV density scaling law (Eq.~14),
the total emission measure values $EM_i = c_3 T_i^4 L_i$ (Eq.~16),
the (hemispheric) flare volumes $V_i = (2 \pi / 3) L_i^3$, and 
the thermal energies $E_{th}= 3 n_i k_B T_i V_i$ (Eq.~17). 
The resulting parameter distributions $N(x)$ for
$x=L_p, T_p, n_p, H_p, EM_p, E_{th}$ are shown in Fig.~8, which
mimic the corresponding observed size distribution closely (Fig.~7).
In the numerical simulations we applied also an emission measure
threshold $EM \ge EM_0=10^{48.5}$ that corresponds to the selection
threshold of GOES M1-class flares in the observations, a maximum
active region size limit $L \le L_{max} \approx 7 \times 10^9$ cm
that corresponds to an upper diameter limit of $2 L_{max} 
\lapprox 140$ Mm for the largest flares observed in this dataset,
and a maximum temperature of $T_{max} \lapprox 20 \times 10^6$ K,
which is a limit imposed by the AIA/SDO EUV temperature filters.
Using these limits, we obtain a dataset of $N(\ge EM_0)=157$ events that
obey the imposed limits, out of 2000 simulated events. 
The observed (Fig.~7) and simulated (Fig.~8)
powerlaw distributions with the slopes $\alpha_L, \alpha_T, \alpha_n,
\alpha_H, \alpha_{EM}$ and $\alpha_{Eth}$ agree with each other 
within the fitting uncertainties in the order of a few percents
(see values of powerlaw slopes indicated in each panel of Fig.~7 and 8).

As mentioned above, the second method to corroborate the RTV scaling  
and the resulting truncation effects theoretically, is by analytical 
calculations of the truncated size distributions, which we derive 
in Appendix A. The final results of these analytically derived
size distributions are shown as red curves in Fig.~8, which 
approximately agree with the Monte-Carlo simulations (black histograms
in Fig.~8) and the observations (Fig.~7). We note that the truncation
effects introduce some slight deviations from strict powerlaws,
whose origin can be fully understood in terms of different truncation
regimes in the analytical calculations (presented in Appendix A).

\subsection{Scaling of Wavelength-Dependent Fluxes}

Finally we want to understand the size distributions $N(F_{\lambda})$
of fluxes $F_{\lambda}$ that are measured with different instrumental
temperature filters, and thus are wavelength-dependent. In Fig.~9
(panels in top half) we show the correlations between the observed 
fluxes $F_{\lambda}$ and the peak emission measure $EM_p$, as they
have been determined from Gaussian DEM fits to the observed fluxes.
In other words, if we know the three physical parameters of a
Gaussian DEM ($EM_p, T_p, \sigma_p$) for a particular flare event,
we want to know how well we can predict the observed fluxes $F_{\lambda}$ with 
a given instrument channel. If a particular wavelength filter has a
broad temperature response or if the observed flare has a broad DEM, 
we would expect a good proportionality between the recorded flux and 
the peak emission measure. The cross-correlation of the fluxes 
$F_{\lambda}$ observed with AIA or GOES and the peak emission measures
$EM_p$ indeed are relatively high in all cases, with cross-correlation
coefficients in the range of
$ccc=0.61-0.86$ (Fig.~9, panels in upper half), and the linear
regression coefficients are in the range of $0.95-1.42$, which
indicates near-proportionality. The best proportionality is found
for the 94, 131, and GOES channels, while the largest deviation 
(with a non-linearity of $F_{304} \propto EM_p^{1.42}$ is found for 
the chromospheric 304 \ang\ channel, as expected.

A more accurate prediction of the observed fluxes $F_{\lambda}$ can
be made by convolving the model (Gaussian) DEM function with the 
instrumental response function $R_{\lambda}(T)$ (Eq.~1), which is shown
in Fig.~9 (panels in lower half). If the data match a Gaussian DEM
perfectly, we expect an exact proportionality between the observed
$F_{\lambda}$ and modeled fluxes $F_{DEM}$. 
The best proportionality is found for 
the 171 \ang\ channel ($F_{171} \propto (F_{DEM})^{1.02}$, $ccc=0.88$), 
the 193 \ang\ channel ($F_{193} \propto (F_{DEM})^{1.10}$, $ccc=0.86$), and 
the 335 \ang\ channel ($F_{335} \propto (F_{DEM})^{1.01}$, $ccc=0.79$),   
while the 304 \ang\ channel shows the largest deviation in terms of
absolute flux. The observed 304 \ang\ flux is almost an order of magnitude
higher than the flare DEM predicts, because this filter is also sensitive
to chromospheric temperatures ($log(T)\approx 4-5$), which is not
included in the single-Gaussian fit of the DEM. A second Gaussian
component would be needed to accomodate a double-peaked DEM function.
However, since we ignored the 304 \ang\ channel in our DEM modeling,
the mismatch of the 304 \ang\ flux is not a problem.  
Although the 304 \ang\ channel is not used in the DEM fit,
its flux can be predicted (Fig.~9) based on the best-fit DEM flux 
from the six coronal channels. The reason why a relatively 
good correlation is found between the observed (mostly chromospheric) 
304 \ang\ flux and the predicted (mostly corona-based) DEM flux is 
probably due to the flare-induced chromospheric heating, which is 
manifested in a commonly visible impulsive brightening of the 304 \ang\ 
flux at beginning of the impulsive flare phase. This good correlation
makes the 304 \ang\ flux to a good flare predictor, which in fact
amounted to 79\% of all flare detections obtained with the two STEREO 
spacecraft (Aschwanden et al.~2013b).

Our DEM-predicted (fitted) fluxes match the observed fluxes with
a mean ratio and standard deviation of $F_{fit}/F_{obs}=1.13\pm0.48$,
which indicates that a single Gaussian fit yields a viable model
for the DEM function of flares.

\section{DISCUSSION}

After we described the data analysis and the theoretical modeling
in the foregoing Sections 2 and 3, we put now the results into a larger 
context, by comparing them with previous studies (Section 4.1, 4.2),
considering implications for the coronal heating problem (Section 4.3),
solar versus stellar scaling laws (Section 4.4), the prediction of the 
largest and smallest solar flare event (Section 4.5), and the prediction 
of powerlaw distribution functions for self-organized criticality models 
(Section 4.6).

\subsection{Tests of the RTV Scaling Law in Previous Studies}

There are very few studies that provide statistics of geometric
flare parameters with simultaneous DEM analysis, as performed here.
In order to provide comparisons with the statistics from previous 
measurements of flare parameters ($A, L, V, D, EM_p, T_p$), we have 
to normalize each of the reported data sets to the same standard 
parameter definitions we are using for AIA data here. This concerns
mostly the definition of the measurement of the flare area $A$, while
the other geometric parameters can be defined in the same standard
way with a circular radius $L=\sqrt{A/\pi}$ and a hemispheric flare
volume $V= (2/3) \pi L^3$. 

The S-054 experiment onboard {\sl Skylab} and the {\sl Solrad 9} satellite
probably provided the first statistics of simultaneous geometric and
emission measure observations of solar flares (Pallavicini et al.~1977). 
In that selection of limb flares, a height $h$ was measured and a flare
volume $V_p$ was estimated (with an unspecified method). From this data set
we estimate the length by $L_p=[(3/2\pi) V_p]^{1/3}$ and the flare area by 
$A_p=\pi L_p^2$. Although the RTV scaling law (Rosner et al.~1978) was not
published yet at that time, we can test it {\sl a posteriori}
based on their tabulated values of volumes $V_p$, peak temperatures $T_p$,
and peak electron densities $n_p$, which is shown in Fig.~(10d). The
RTV-predicted emission measure $EM_{RTV}=c_4 T_p^3 L_p^2$ (Eq.~20)
agrees remarkably well with the measured values $EM_p$, within a factor
of $(EM_{RTV}/EM_p)=10^{0.51\pm0.56} \approx 3.2$ (Fig.~10d), similar to
our study with $(EM_{RTV}/EM_p)=10^{0.44\pm0.50} \approx 2.7$ (Fig.~10f).
So, our AIA/SDO results are perfectly consistent with the flare statistics
obtained from {\sl Skylab} and {\sl Solrad 8} in the 1-8 \ang\ spectral range.

Statistics of (non-imaging) flare peak emission measures $EM_p$ and
temperatures $T_p$ was obtained for 540 flares observed with Yohkoh/BCS
(using the Fe XXV line) and GOES, covering a range of $EM_p \approx
10^{47}-10^{49}$ cm$^{-3}$ and $T_p \approx 12-25$ MK (Feldman et al.~1995a).
These values were measured at the flare peak time $t_p$, which appears to
correspond to the time of the GOES peak flux. A follow-up study extended
the ranges to $EM_p \approx 10^{47}-10^{51}$ cm$^{-3}$ and $T_p \approx 
5-40$ MK (Feldman et al.~1995b), as well as an additional study including 
flares from A2 to X2 GOES class (Feldman et al.~1996). The total emission
measures $EM_p$ are found to coincide with the datapoints from AIA/SDO. 

The relationship of the the RTV-predicted loop length $L_{RTV}=
c_3 T_e^2 n_e^{-1}$ (Eq.~15) with the actually measured loop length
$L$ was investigated for 32 resolved loops observed with Yohkoh/SXT
(Kano and Tsuneta 1995), and a deviation from the RTV law was claimed.
However, since the range of loop lengths $L$ extends over less than a
decade and the statistical sample is small, we do not consider their 
result as significant. Their plot of $L_{RTV}$ versus $L$ (Fig.~9
in Kano and Tsuneta 1996) ressembles our scatterplot from AIA/SDO 
for flare events (Fig.~6c), which agrees with the RTV prediction
within a factor of $L_{RTV}/L_p=1.3\pm2.1$.

Imaging observations from Yohkoh/SXT were used to measure the loop half
length $L_{half}$, loop apex temperature $T_{max}$, flare rise $\tau_r$ 
and decay time 
$\tau_d$ of 19 flare loops (Metcalf and Fisher 1996), and were combined
with emission measure $EM_p$ and temperature $T_p$ measurements from
GOES (Garcia 1998). We plot the RTV-predicted flare peak emission
measures $EM_{RTV}$ versus the observed values $EM_p$ for the 14 flares
of Garcia (1998) in Fig.~(10e) and find a trend of over-prediction by
a factor of ($EM_{RTV}/EM_p)=10^{1.76\pm0.63}\approx 60$), which probably
results from the unreliability of estimating loop lengths $L_p$ from flare
decay times $\tau_d$ (see also discussions in Metcalf and Fisher 1996;
Hawley et al.~1995; G\"udel 2004; Reale 2007), as well as from over-estimates
of the peak temperature $T_p$ (which scales with the fourth power in the
emission measure, Eq.~16). 

Flare peak emission measures span over a huge range, from $EM_p \lapprox
10^{51}$ cm$^{-3}$ for the largest (solar) GOES X-class flares down by
8 orders of magnitude to $EM_p \gapprox 10^{43}$ cm$^{-3}$ 
for the smallest nanoflares detected in EUV. Such nanoflare statistics
exists for 23 Quiet-Sun brightening events observed with
SohO/EIT (Krucker and Benz 2000) and for nanoflares observed with TRACE 
(Aschwanden et al.~2000). Active region brightenings were observed
with slightly higher emission measure in the range of
$EM_p \approx 10^{44}-10^{48}$ cm$^{-3}$ with Yohkoh/SXT (Shimizu 1995).
We perform a test of the RTV-predicted emission measure $EM_{TRV}$
versus the observed emission measure $EM_p$ and find some significant
over-estimation by the RTV scaling law, i.e., 
($EM_{RTV}/EM_p)=10^{0.84\pm0.19}\approx 7$; Fig.~10a, 
Krucker and Benz 2000), 
($EM_{RTV}/EM_p)=10^{0.28\pm0.33}\approx 20$; Fig.~10b, 
Aschwanden et al.~2000), and  
($EM_{RTV}/EM_p)=10^{1.84\pm0.69}\approx 70$; Fig.~10c, 
Shimizu 1995). 
Since the emission measure scales with the fourth power of the
temperature ($EM_p=c_4 T_p^4 L_p$, Eq.~16), it is conceivable that
the temperature $T_p$ is over-estimated for these events. Hydrodynamic
simulations show a drop of the temperature maximum $T_{max}$ during a 
flare by a factor of $\approx 2$ to the temperature $T_p$ when the
flare reaches the peak emission measure $EM_p$ or peak density $n_p$
(Fig.~5; Aschwanden and Tsiklauri 2009). 
Thus, if the flare maximum temperature $T_m$ is used in the RTV scaling
law, instead of the cooler temperature $T_p$ during the emission measure
peak, the RTV-predicted emission measure could be over-estimated by a 
factor of up to $(T_m/T_p)^4 \approx 2^4 = 16$. Careful evolutionary
flare studies with determination of the time profiles $T_e(t)$ and
$n_e(t)$ (as shown in Fig.~1) are needed to bring clarity into this 
question.

\subsection{Flare Decay Time Scaling Laws}

While the RTV law applies only to coronal loops that have a balance 
between the volumetric heating rate and the (radiative and conductive) 
loss rate, or to dynamic flare loops at the equilibrium point of 
energy balance, alternative scaling laws have been derived for
other dynamic phases during flares, such as during the decay phase
(Serio et al.~1991). At the beginning of the flare decay phase, a
thermodynamic decay time $\tau$ with the scaling law
$\tau \propto L T^{-1/2}$ was derived (Serio et al.~1991), 
but was found to lead to a moderate overestiate of the loop length 
(Reale 2007). The temperature-density relationship was found to scale 
as $T_e \propto n_e^2$ during the (dynamic) decay phase of the flare 
(Jakimiec et al.~1992), rather than $T_p \propto n_p^{1/2} L_p^{1/2}$ as
expected for the RTV law (Eq.~13) in the case of steady-state heating.
This scaling law was found to apply to solar flare observations
(Sylwester et al.~1993; Metcalf and Fisher 1996; Bowen et al.~2013) 
as well as to 
stellar flare observations (e.g., see review by G\"udel 2004). 
Since the e-folding flare decay time $\tau$ is an additional 
free parameter or observable that is not measured here, it has no
consequence on our derived relationships, but is consistent with
the temperature-density phase diagram obtained from hydrodynamic
simulations (Jakimiec et al.~1992; Aschwanden and Tsiklauri 2009),
as shown in Fig.~1 (right panel), and corroborates the restriction
that the RTV law can only be applied to the instant of energy balance 
at the flare peak time, but not later on during the decay phase.

\subsection{Implications for the Coronal Heating Problem}

It was already pointed out early on that powerlaw distributions
$N(E)$ of energies with a slope flatter than the critical value of
$\alpha_E=2$ imply that the energy integral diverges at the upper end,
and thus the total energy of the distribution is dominated by the
largest events (Hudson et al.~1991). On the opposite side, if
the powerlaw distribution is steeper than the critical value, it will
diverge at the lower end, and thus the total energy budget will
be dominated by the smallest detected events, an argument that was 
used for dominant nanoflare heating in some cases with insufficient
wavelength coverage of solar nanoflare statistics (e.g., 
Krucker and Benz 2000). The powerlaw slope $\alpha_E$ for energies
depends sensitively on its definition (e.g., Benz and Krucker 2002),
in particular on the assumptions
of the flare volume scaling $V(A)$ that has to be inferred from 
measured flare areas $A$ in the case of thermal energies,
$E_{th}=3 n_e k_B T_e V$ (Eq.~17). In the present study, where
we selected only large flares (of M and X GOES class), we found
a powerlaw slope of $\alpha_{Eth}=1.66\pm0.13$ for the thermal
energies $E_{th}$, which closely matches the powerlaw distributions
of non-thermal energies determined from hard X-ray producing
electrons, e.g.,
$\alpha_{nth}=1.53\pm0.02$ for a much larger sample including
smaller flares (Crosby et al.~1993). Thus, based on the statistics
of large flares we do not see any evidence that would support
nanoflare heating, at least not for flares with energies $\gapprox
10^{29}$ erg. This argument, however, does not rule out that the
powerlaw slope could steepen at smaller energies. Synthesized
flare energy statistics on all scales (e.g., Fig.~10 in Aschwanden
et al.~2000) are composed of measurements with different event
selection criteria, different detection methods, and different
energy definitions. In future studies we plan to extend the current
flare statistics with the same method to smaller energies below
$10^{29}$ erg, in order to obtain a self-consistent flare energy
distribution on all scales. 

\subsection{Solar versus Stellar Flare Scaling Laws}

A comparison of solar and stellar flare scaling laws has been
compiled in Aschwanden, Stern, and G\"udel (2008), where an
empirical scaling of the peak emission measure $EM_p$ with peak
temperature $T_p$ of $EM_p \propto T_p^{4.7}$ was found for both
populations, but the stellar flares exhibit about a factor of
$\approx 250$ times higher emission measures than the solar flares
(at the same flare peak temperature). We present a similar 
scatterplot in Fig.~11, where we sbow the new measurements
of the 155 M and X-class flares obtained from AIA/SDO and GOES.
In addition we overlay the predicted relationship for the RTV
law for both constant loop lengths and constant heating rates.
For constant loop lengths, the predicted RTV scaling is
$EM_p \propto T_p^4$ (Eq.~17), while a constant heating rate
implies $L_p \propto T_p^{7/4}$ (using Eq.~18), which yields
$EM_p \propto T_p^{5.75}$ (by inserting Eq.~18 into Eq.~17).
An alternative relationship of $EM_p \propto T_p^{8.5}$ was
derived for a magnetic reconnection model (Shibata and Yokoyama 1999).

Following the RTV predictions we can see from Fig.~11
that most of the large solar flares are produced with heating
rates of $H \approx 10^{-2},...,1$ erg cm$^{-3}$ s$^{-1}$
on spatial scales of $L_p \approx 10^8-10^{10}$ Mm.
Stellar flares exhibit a higher range of heating rates
and also require larger spatial scales of $L \approx 10^9-10^{12}$ cm.
It appears that the largest stellar flares occupy larger volumes
than solar flares, while the smallest detected stellar flares 
have similar sizes as the largest solar flares.

\subsection{Predicting the Largest and Smallest Solar Flare}

The scaling laws we established here set a firm upper limit on
the largest flare events. The temperature distribution $N(T_p)$
drops off steeply at $T_e \approx 20$ MK (Fig.~2d), so that this 
value can be considered as an upper limit on temperatures $T_p$ as 
measured at the peak of the DEM distribution function. 
Spatial scales are found to have a cutoff at $L_p \lapprox 7 \times
10^9$ cm, corresponding to a diameter of $2L_p \lapprox 1.4 \times
10^{10}$ cm (or 140 Mm), which corresponds to 10\% of the solar
diameter, which is physically dictated by the maximum size of the
largest active regions. The maximum predicted thermal energy in
solar flares is thus, $E_{th} = c_5 T_p^4 L_p^2 \lapprox 3 \times
10^{32}$ erg (Eq.~17).  

This energy estimate of the largest solar flare possible agrees with
a recent study, which states that the largest solar flares observed
over the past few decades have reached energies of a few times
$10^{32}$ ergs (Aulanier et al.~2013). Alternatively, the same
authors estimated the largest amount of released magnetic energy
possible from assuming a flare size that covers 30\% of the largest
sunpspot group ever reported, with its peak magnetic field being set
to the strongest value ever measured in a sunspot, which then 
produces a flare with a maximum magnetic energy of $\approx 6 \times
10^{33}$ ergs, which is a factor of 20 higher than the upper limit
of thermal energies based on 
the RTV scaling. However, only a fraction of the magnetic energy
is converted into thermal energy, a factor that is estimated to be
in the order of $\approx 1\%-10\%$ (Emslie et al.~2004, 2005, 2013). 

In the other extreme, we can also predict the energy of the smallest 
flare using the relationship of Eq.~(17). A lower limit for the
temperature is given by the background corona, which is approximately
$T_{min}=1.0$ MK in active regions. The minimum length scale is given
by the smallest loop segment that sticks out of the transition region, 
which we estimate to $L_{min}\approx 1.0$ Mm, given the chromospheric
height of $h \approx 2.0$ Mm. Thus the smallest detectable nanoflare
is expected to have a thermal energy of 
$E_{th} = c_5 T_p^4 L_p^2 \gapprox 7 \times 10^{24}$ erg (Eq.~17),
which is almost 8 orders of magnitude smaller than the largest
predicted flare and justifies the term {\sl nanoflare}.
Since the dataset analyzed here includes only large flares,
extrapolations of the scaling laws to the nanoflare regime may bear
large uncertainties, which will be reduced in future studies that
include smaller flare events.

\subsection{Predicting Powerlaw Distributions for SOC Models}

Self-organized criticality (SOC) models can be characterized by
powerlaw-like occurrence frequency or size probability distribution
functions. After we investigated various scaling laws between 
observables and inferred physical parameters, the question arises now
whether we are able to predict all observed distribution functions 
with the correct powerlaw slopes from first principles. 
A diagram of the application of the fractal-diffusive 
(FD-SOC) model to the physical system of solar flares is provided in
Fig.~12, visualized as a flow chart that progresses from the input
parameters (left) to the output distribution functions (right) in
three different regimes: (i) The spatio-temporal regime that follows
from universal statistics (Fig.~12, top part), (ii) the regime of the
physical system that is described by the hydrodynamic RTV scaling laws 
(Fig.~12, middle part), and (iii) the observers regime that is
characterized by the response functions and detection sensitivity
of a specific instrument (Fig.~12, bottom). From this diagram 
we can easily see what input parameters are required to predict all
output distribution functions: the Euclidean dimension $d$ of the system,
limits on the minimum and maximum length scales ($L_{min}, L_{max}$),
the lower limit $H_0$ of the heating rate that triggers a flare
instability, the powerlaw index $\alpha_H$ of the heating
function, the instrumental response functions $R_{\lambda}(T)$, and
the flux threshold $F_{thresh}$ or emission measure threshold $EM_0$
of the flare event detection algorithm, the diffusion coefficient
$\kappa$, the diffusion spreading exponent $\beta$, and the Gaussian
width $\sigma_p$ of the DEM distribution function, which all have
been determined in the present analysis for a representative sample 
of large (M and X GOES class) flares. The best-fitting values were
found to be: $d=3$, $L_{min}=4$ Mm, $L_{max} =70$ Mm, $H_0=0.04$
erg cm$^{-2}$ s$^{-1}$, $\alpha_H=1.8$, $EM_0=10^{48.5}$ (for GOES
M1.0 class level), $\beta=0.53\pm0.27$, $\kappa \approx 52$ 
km s$^{-\beta/2}$, $\sigma_p=0.50\pm0.13$. Using these input
parameters, we can predict every probability distribution function
$N(x)$ for the parameters $x=(L, A, V, \tau, T_p, n_p, E_{th}, EM_p)$. 
In addition, choosing a suitable instrument sensitive to both EUV
and soft X-ray wavelengths, we need in addition to know the instrumental
response functions $R_{\lambda}(T)$ as a function of the temperature $T$,
and can then predict the probability distribution functions $N(F_{\lambda})$
of the flux $F_{\lambda}$ for any arbitrary wavelength channel $\lambda$.

Ultimately, this framework could be developed further, by including
scaling laws and distribution functions of the magnetic field $B$,
since we expect that the primary energy source of the heating process
comes from dissipation of magnetic energy. This requires the knowledge
of a scaling law between the heating rate $H(B, L_p)$ and the magnetic field
strength $B$ and length scales $L_p$, as tentatively discussed in 
Appendix B and to be examined in a future study.

\section{CONCLUSIONS}

While we analyzed the spatio-temporal evolution of a complete set of
155 flares (above the M and X-GOES class level) in Paper I, 
we conduct a differential emission measure
(DEM) analysis on the same set of flares in this Paper II, in order to derive
statistics, size distributions, and scaling laws of physical parameters measured 
during the peak times of the flares, such as electron temperatures, electron 
densities, emission measures, and thermal energies. The DEM analysis is based 
on the six coronal wavelength filters of AIA (94, 131, 171, 193, 211, 335),
which allow us temperature diagnostics in the range of $T_e \approx 1-20$ MK.
The quality of the DEM results measured at the flare peak time $t_p$ carried out 
here is comprehensive in wavelength and time coverage,
since we carry out DEM fits throughout the flare impulsive 
phase (in order to detect the emission measure peak time $t_p$) and since AIA 
provides sensitivity in a sufficiently broad temperature range to reliably detect 
the peak temperature $T_p$ at the spectral peak $EM_p$ of the DEM.
The major conclusions of this statistical study are:

\begin{enumerate}

\item{The six observed fluxes $F_{\lambda}$ of the coronal wavelength filters
	of AIA/SDO measured during the soft X-ray peak times $t_p$ of
	virtually all analyzed flares can be fitted with a
	single-Gaussian {\sl differential emission measure (DEM)} function,
	with an average goodness-of-fit of $F_{fit}/F_{obs}=1.13\pm0.48$.
	These DEM fits provide the three key parameters of the flare 
	peak emission
	measure $EM_p$, peak temperature $T_p$, and Gaussian temperature width
	$\sigma_p$ of the DEM. These parameters were determined in the ranges 
	of $log(EM_p)=47.0-50.5$, $T_p \approx 5.0-17.8$ MK, and $\sigma_p
	=0.22-0.75$ (or $\sigma_p=0.50\pm0.14$).}

\item{Measuring the time-integrated flare area $A=\pi L_p^2$ with radius $L_p$, 
	assuming a hemispheric flare volume $V=(2/3) \pi L_p^3$, and a filling
	factor of unity, we derive the average electron density $n_p=\sqrt{(EM_p/V)}$
	and the thermal energy $E_{th}=3 n_p k_B T_p V$ at the flare peak time,
	which are found in the ranges of $n_p=3.8 \times 10^9 - 8.8 \times 10^{11}$
	cm$^{-3}$ and $E_{th}=1.6 \times 10^{28} - 1.1 \times 10^{32}$ erg.}

\item{Using the parameters $L_p$, $T_p$, and $n_p$ we test the RTV scaling law and 
	find an excellent agreement between the RTV-predicted and observed parameters,
	with a mean of $T_{RTV}/T_{obs}=1.05\pm0.38$ or 
	$L_{RTV}/L_{obs}=n_{RTV}/n_{obs}=1.3\pm2.1$. This agreement implies that
	energy balance between the heating rate and radiative and conductive loss rates 
	is achieved during the flare peak time $t_p$, which permits the applicability
	of the RTV scaling law near this particular time in the flare evolution, 
	although the hydrodynamic evolution is not stationary and the heating is
	not spatially uniform, as assumed in the original derivation of the RTV scaling law.}

\item{The RTV scaling laws allow us to calculate the probability distributions
	of the flare peak temperatures, $N(T_p) \propto T_p^{\alpha_T}$, 
	peak electron densities, $N(n_e) \propto n_p^{\alpha_n}$, peak emission measures 
	$N(EM_p) \propto EM_p^{\alpha_{EM}}$, and thermal energies $N(E_{th}) \propto 
	E_{th}^{\alpha_{Eth}}$, if the
	size distributions of length scales $N(L_p)$ and heating rates $N(H)$ are known.
	The length scale distribution, postulated by the scale-free probability conjecture,
	$N(L_p) \propto L_p^{-3}$, is found to be consistent with the observations. 
	Assuming a powerlaw function for the heating rates, $N(H) \propto H^{-\alpha_H}$,
	with a slope of $\alpha_H=1.8$ and a cutoff at $H \gapprox H_0=0.04$ erg cm$^{-3}$ 
	s$^{-1}$ yields an accurate match for the observed powerlaw slopes within the
	uncertainties of the powerlaw fits, i.e., $\alpha_{EM}=1.78\pm0.03$, 
	$\alpha_n=2.15\pm0.17$, and $\alpha_{Eth}=1.66\pm0.13$.}

\item{We show also how the size distributions of $T_p$, $n_p$, $EM_p$, $E_{th}$ can be
	calculated analytically, using the RTV scaling laws, the truncation effects
	due to the emission measure or flux threshold, the truncation effects
	caused by lower heating rate limit, and the lower and upper length scale limit.
	Using the instrumental response functions $R_{\lambda}(T)$ of a wavelength
	filter $\lambda$, we can also model the observed fluxes $F_{\lambda}$ in each
	wavelength $\lambda$, their size distributions (which all turn out to be
	$N(F_{\lambda}) \propto F_{\lambda}^{-2}$), and the high degree of correlations
	in flux-flux, flux-volume, or flux-emission measure relationships.}

\item{Our result of the inferred volumetric heating rate size distribution 
	$N(H) \propto H^{-1.8}$ predicts identical size distributions
	for the magnetic field, $N(B) \propto B^{-1.8}$, and the magnetic fluxes, 
	$N(\Phi) \propto \Phi^{-1.8}$, and is consistent with the
	statistics of magnetic fluxes on the solar surface measured by
	Parnell et al.~(2009), as well as with the heating flux scaling law
	$F_H = H \times L \propto B L^{-1}$ found from hydrostatic simulations 
	of the entire Sun's corona by Schrijver et al.~(2004).}

\item{The size distribution of thermal flare energies is found to be 
	$N(E_{th}) \propto E_{th}^{1.66\pm0.13}$, which is close to the
	size distribution of non-thermal flare energies calculated from
	hard X-ray producing electrons, $N(E_{nth}) \propto E_{nth}^{1.53\pm0.02}$.
	This finding of a powerlaw slope below the critical value of 2 corroborates
	that the energy dissipated in the solar corona is dominated by the largest
	flares. If this distribution holds down to the smallest energies (which is
	the subject of future studies), heating of the solar corona by nanoflares 
	can be ruled out. The dataset analyzed here, however, includes only large flares,
	and thus no conclusion can be drawn about the significance of nanoflares.}

\item{The RTV scaling law allows us to predict the largest and smallest solar flare,
	which is estimated to be $E_{th,min} \gapprox 7 \times 10^{24}$ erg for the smallest
	flare, and $E_{th,max} \lapprox 3 \times 10^{32}$ erg for the largest flare,
	spanning a range of almost 8 orders of magnitude in energy. For the analyzed
	sample of M and X-class flares we find an energy range of 3 orders of
	magnitude ($E_{th} \approx 10^{29}-10^{32}$ erg)}.

\item{Comparing solar with stellar flares and applying the RTV predictions 
	we find that most of the large solar flares are produced with heating
	rates of $H \approx 10^{-2},...,10^{2}$ erg cm$^{-3}$ s$^{-1}$
	on spatial scales of $L_p \approx 10^8-10^{10}$ Mm.
	Stellar flares exhibit a similar range of heating rates
	but require larger spatial scales of $L \approx 10^9-10^{12}$ cm.
	It appears that the largest stellar flares occupy larger volumes
	than solar flares, while the smallest detected stellar flares 
	have similar sizes as the largest solar flares.}

\end{enumerate}

This study conveyed deeper physical insights into nonlinear phenomena controlled
by self-organized criticality. We have shown that the nonlinear nature of
scaling laws generally predicts powerlaw distribution functions for most
observables, especially for energy distribution functions, a key parameter to
characterize the size of SOC avalanches. Our concept of the fractal-diffusive
SOC model (Aschwanden 2012a) provides a framework to relate powerlaw distribution
functions and correlations of SOC parameters to physical scaling laws that govern
SOC phenomena. For solar flares we found that the hydrodynamic processes can be
formulated with the RTV scaling law during the flare peak time, what allowed us
to retrieve the size distribution of average heating rates in flares, which turned 
out to be identical to the magnetic flux distribution (as measured by
Parnell et al.~2009). Future studies with measurements of the magnetic field 
in flare sites are likely to reveal us scaling laws between the 
magnetic energy dissipation rate and plasma heating in solar flares.

\acknowledgements
The author acknowledges helpful comments of anonymous referees,
discussions, and software support
of the AIA/SDO team. This work has benefitted from fruitful
discussions with Karel Schrijver, Henrik Jensen, Nicholas Watkins, 
J\"urgen Kurths,
and by the {\sl International Space Science Institute (ISSI)}
at Bern Switzerland, which hosted and supported a workshop on
{\sl Self-Organized Criticality and Turbulence} during October
15-19, 2012. This work was partially supported by NASA contract
NNX11A099G ``Self-organized criticality in solar physics''
and NASA contract NNG04EA00C of the SDO/AIA instrument to LMSAL.

\subsection*{Appendix A: Analytical Calculation of Truncation Effects 
	in Powerlaw Size Distributions using the RTV Law}

In this Appendix we calculate quantitatively how truncation effects
and incomplete sampling affects the observed powerlaw distributions.

In our study we sampled only 
flares larger than the GOES M1.0 class, which represents a lower limit 
of the flux or emission measure. This is also true for most other datasets, 
since event catalogs are generally compiled with some completeness down 
to an instrument-dependent flux threshold or a related selection criterion.
Thus, we have to calculate how a flux or emission measure threshold,
$EM \ge EM_0$, scales for each parameter. From Eq.~(16) we obtain directly
how the length limit $L_1$ scales for a fixed threshold
value $EM_0$ as a function of the temperature $T_p$,
$$
	L_1(T_p) = \left({ EM_0 \over c_4} \right) T_p^{-4} \ ,
	\eqno(A1)
$$
or on the electron density $n_p$,
$$
	L_1(n_p) = \left({3 \over 2 \pi} EM_0 \right)^{1/3}
		 n_p^{-2/3} \ .
	\eqno(A2)
$$
Similarly, by inserting the thermal energy $E_{th}$ from the RTV
scaling law (Eq.~17) we obtain,
$$
	L_1(E_{th}) = \left( c_5^{4/3} {EM_0 \over c_4}\right)^{-3/5}
		E_{th}^{4/5} \ ,
	\eqno(A3)
$$
or for the heating rate $H$ (Eq.~18),
$$
	L_1(H) = \left( {c_6^{8/7} EM_0 \over c_4}\right)^{7/23}
		H^{-8/23} \ .
	\eqno(A4)
$$
These truncation boundaries are overplotted onto the scatterplots
of the observables $L_p$ versus $T_p$, $n_p$, $H_p$, $EM_p$, and
$E_{th}$ in Fig.~7 (dashed lines). We see that these
truncation boundaries constitute a lower limit of length
scales $L_1$ for the parameters $T_p$, $n_p$, and $H_p$, but
an upper limit of length scales $L_1$ for the thermal energies $E_{th}$.
These length scale limits $L_1$ quantify the undersampling and
threshold effects due to a lower emission measure limit or
instrumental flux detection threshold, which we have to implement in the
analytical derivation of the size distributions and powerlaw slopes.

\bigskip
The lower cutoff $H_0$ of the heating rate distribution causes also
truncation effects, as it can be seen in the scatterplots in Fig.~7
(dotted linestyle). We calculate how the heating rate cutoff
$H \ge H_0$ scales with each parameter. From Eq.~(18) we obtain directly
how the length limit $L_2$ scales for a fixed heating rate 
value $H_0$ as a function of the temperature $T_p$,
$$
	L_2(T_p) = \left({ c_6 \over H_0} \right)^{1/2} T_p^{7/4} \ ,
	\eqno(A5)
$$
or on the electron density $n_p$ (using Eqs.~18 and 14),
$$
	L_2(n_p) = \left( {c_6 c_1^{7/2} \over H_0} \right)^{4}
		 n_p^7 \ .
	\eqno(A6)
$$
Similarly, by inserting the peak emission measure $EM_p$ from the RTV
scaling law (Eqs.~16 and 18) we obtain,
$$
	L_2(EM_p) = \left( {c_6 \over H_0 c_4^{7/8}}\right)^{8/23}
		EM_p^{7/23} \ ,
	\eqno(A7)
$$
or for the thermal energy $H$ (Eqs.~17 and 18),
$$
	L_2(E_{th}) = \left( {c_6 \over H_0 c_5^{7/6}}\right)^{3/13}
		E_{th}^{7/26} \ .
	\eqno(A8)
$$
These truncation boundaries are overplotted onto the scatterplots
of the observables $L_p$ versus $T_p$, $n_p$, $H_p$, $EM_p$, and
$E_{th}$ in Fig.~7 (dotted lines). We see that these
truncation boundaries constitute an upper limit of length
scales $L_2$ for the parameters $T_p$, $n_p$, $EM_p$, and $E_{th}$.
These length scale limits $L_2$ affect the size distributions 
in those regimes where $L_2 < L_{max}$, which we have to consider 
in the calculation of the size distributions in the next subsection.

\bigskip
Using the two size distributions $N(L_p)$ (Eq.~19) and $N(H)$ (Eq.~20), 
we can calculate analytically the predicted size distributions $N(x) dx$ for 
all physical parameters $x=T_p, n_p, EM_p, E_{th}, H_p$,
by integrating the length scale distribution $N(L)$ from
the minimum $L_1(x)$ to the maximum value $L_2(x)$, and by substituting 
the parameters $x$ according to the RTV scaling law relationships given 
in Eqs.~(13-18), 
$$
   N(x) = \int_{L_1(x)}^{L_2(x)} N(L,H,x) \ dL 
	= \int_{L_1(x)}^{L_2(x)} N(L) \ N(H[L,x]) \ dL 
	= \int_{L_1(x)}^{L_2(x)} N_0 \ L^{-3} H[L,x]^{-\alpha_H} \ dL \ ,
	\eqno(A9)
$$
where the integration boundaries $L_1(x)$ and $L_2(x)$ have to be
adjusted to the cutoffs $L_1(x)$ (Eqs.~A1-A4) imposed by the 
emission measure threshold $EM_0$ and the cutoffs $L_2(x)$ set by the
minimum heating rate $H_0$ (Eqs.~A5-A8).

We start with the occurrence frequency of temperatures $N(T_p)$,
for which we obtain,  
$$
  N(T_p) \propto \int_{L_1(T_p)}^{L_2(T_p)} 
                 T^{-(7/2)\alpha_H} \ L_p^{-3+2 \alpha_H} \ dL 
         \propto T^{-(7/2)\alpha_H} \  
	 \left[-L_2(T_p)^{2(\alpha_H-1)}+L_1(T_p)^{2(\alpha_H-1)} \right] \ ,
	\eqno(A10)
$$
where the minimum temperature $T_{min}$ is defined by the intersection point
of the emission measure threshold $L_1(T_p)$ (Eq.~A1) with the heating 
threshold $L_1(H)$ (Eq.~A4), at
$$
	T_p \ge T_{min}=\left[ \left( {EM_0 \over c_4} \right)
		\left( {H_0 \over c_6} \right)^{1/2} \right]^{4/23} \ ,
	\eqno(A11)
$$
which amounts to $T_{min}=5.25$ MK for our model with $EM_0=10^{48.5}$
cm$^{-3}$ and $H_0=0.04$ erg cm$^{-3}$. The lower integration limit
is given either by the emission measure threshold limit at $L_1(T_p, EM_0)$ 
(Eq.~A1) or the length scale minimum $L_{min}$, while the upper integration 
limit is given by the heating rate limit $L_2(T_p, H_0)$ (Eq.~A5) or
the length scale maximum $L_{max}$, which are all indicated in the
Figs.~6 and 8 (panels b). The so obtained analytical function of the 
temperature distribution $N(T_p)$ is shown in Fig.~9b (red curve),
which is confined to a narrow range that increases from $T_{min}=5.25$ MK 
steeply to a maximum at $T_e=8.85$ MK, and drops again steeply in the 
range of $T_e=8.85-20$ MK to the sensitivity limit of the AIA instrument. 

\medskip
The second distribution we are going to calculate is for the electron
peak density, $N(n_p)$. Again, by inserting the heating rate
relationship $H(T_p, L_p)$ (Eq.~18) and the RTV relationship
$T_p \propto (n_p L_p)^{1/2}$ (Eq.~13) into the general distribution 
$N(x)$ (Eq.~A9) with $x=n_p$, we predict the following density 
distribution $N(n_p)$, 
$$
  N(n_p) \propto \int_{L_1(n_p)}^{L_2(n_p)} 
                 n_p^{-(7/4)\alpha_H} \ L_p^{-3+ \alpha_H/4} \ dL 
         \propto n_p^{-(7/4)\alpha_H} \  
	 \left[-L_2(n_p)^{-2+\alpha_H/4}+L_1(T_p)^{-2+\alpha_H/4} \right] \ ,
	\eqno(A12)
$$
where the minimum density $n_{min}$ is defined by the intersection point
of the emission measure threshold $L_1(n_p)$ (Eq.~A2) with the heating 
threshold $L_1(n_p)$ (Eq.~A2), at
$$
	n_p \ge n_{min}=\left[ \left( {{3 \over 2 \pi} EM_0 } \right)^{1/3}
		\left( {H_0 \over c_6 c_1^{7/2}} \right)^{4} \right]^{3/16} \ ,
	\eqno(A13)
$$
which amounts to $n_{min}=8.25 \times 10^9$ cm$^{-3}$ for our model with 
$EM_0=10^{48.5}$ cm$^{-3}$ and $H_0=0.04$ erg cm$^{-3}$. The lower integration 
limit is given either by the emission measure threshold limit at 
$L_1(n_p, EM_0)$ (Eq.~A2) or the length scale minimum $L_{min}$, while the 
upper integration limit is given by the heating rate limit $L_2(n_p, H_0)$ 
(Eq.~A6) or the length scale maximum $L_{max}$, which are all indicated in the
Figs.~7c and 8c. The so obtained analytical function of the 
density distribution $N(n_p)$ is shown in Fig.~9c (red curve),

\medskip
The third distribution we are calculating is the emission measure
distribution $N(EM)$, by inserting the RTV relationships 
for $H(T_p, L_p)$ (Eq.~18) and $T_p(EM_p, L_p)$ (Eq.~13) into the
general distribution $N(x)$ (Eq.~A9) with $x=EM_p$, 
$$
  N(EM_p) \propto \int_{L_1(EM_p)}^{L_2(EM_p)} 
                 EM_p^{-(7/8)\alpha_H} \ L_p^{-3+ (23/8)\alpha_H} \ dL 
         \propto EM_p^{-(7/8)\alpha_H} \  
	 \left[-L_2(EM_p)^{-2+(23/8)\alpha_H}+L_{min}^{-2+(23/8)\alpha_H} 
	\right] \ ,
	\eqno(A14)
$$
where the lower integration limit is given either by the 
length scale minimum $L_{min}$, while the upper integration limit is 
given by the heating rate limit $L_2(EM_p, H_0)$ (Eq.~A7) or the length 
scale maximum $L_{max}$, which are all indicated in the
Figs.~7e and 8e. The so obtained analytical function of the 
emission measure distribution $N(EM_p)$ is shown in Fig.~9e (red curve),
which represents a powerlaw function with a slope of 
$\alpha_{EM}=1.73\pm0.07$.

\medskip
Similarly we calculate the distribution of thermal energies $N(E_{th})$,
by inserting the RTV relationships for $H(T_p, L_p)$ 
(Eq.~18) and $T_p(E_{th}, L_p)$ (Eq.~17) into the general distribution 
$N(x)$ (Eq.~A9) with $x=E_{th}$ and predict the 
following distribution for $N(E_{th})$,
$$
  N(E_{th}) \propto \int_{L_1(E_{th})}^{L_2(E_{th})} 
                 E_{th}^{-(7/6)\alpha_H} \ L_p^{-3+ (13/3)\alpha_H} \ dL 
         \propto E_{th}^{-(7/6)\alpha_H} \  
	 \left[-L_2(E_{th})^{-2+(13/3)\alpha_H}+L_{min}^{-2+(13/3)\alpha_H} 
	\right] \ ,
	\eqno(15)
$$
where the lower integration limit is given by the 
length scale minimum $L_{min}$, while the upper integration limit is 
given by the emission measure threshold at $L_1(E_{th})$ (Eq.~A3), 
the heating rate limit $L_2(E_{th}, H_0)$ (Eq.~A8), or the length 
scale maximum $L_{max}$, which are all indicated in the
Figs.~7f and 8f. The so obtained analytical function of the 
emission measure distribution $N(E_{th})$ is shown in Fig.~8f (red curve),
which represents a powerlaw function with a slope of 
$\alpha_{EM}=1.64\pm0.06$.

\medskip
The analytically calculated size distributions (shown in Fig.~8, red
curves) exhibit approximate powerlaw functions, as well as slight changes
in the slopes and turnover points, all caused by truncations in
the datapoints due to the emission measure threshold $EM_0$ and
the heating rate limit $H_0$. For sake of simplicity we used an 
exact powerlaw distribution of length scales $N(L) \propto L^{-3}$ in
the analytical calculations, which neglects truncation effects 
for small length scales $L$, and thus shows some deviations from
the numerically simulated distributions in Fig.~8. More accurate
size distributions are obtained with the Monte-Carlo simulations 
shown in Fig.~8, which compares favorably with the observed values shown 
in Fig.~7. In particular we match the powerlaw slopes of heating rates 
($\alpha_H^{obs}=1.45\pm0.05$ versus $\alpha_H^{sim}=1.47\pm0.09$), 
for peak emission measures,
($\alpha_{EM}^{obs}=1.78\pm0.03$ versus $\alpha_{EM}^{sim}=1.73\pm0.07$), 
and for thermal energies,
($\alpha_{Eth}^{obs}=1.66\pm0.13$ versus $\alpha_{Eth}^{sim}=1.64\pm0.06$). 
The satisfactory match within the stated uncertainties corroborates the
validity of our numerical and analytical models.

\subsection*{Appendix B: Scaling Law of the Coronal Heating Rate}

For the heating rate distribution $N(H) \propto H^{-\alpha_H}$ 
(Eq.~20) we made the assumption of a powerlaw function with an
unknown powerlaw exponent $\alpha_H$, which turned out to require
a value of $\alpha_H=1.8$ in Monte-Carlo simulations in order to
reproduce the observed distributions of $\alpha_H^{obs}=1.45$,
$\alpha_n=2.15$, $\alpha_{EM}=1.78$, and $\alpha_{Eth}=1.66$
(Fig.~7c-f). Can we explain this particular value with a physical
model from first principles? 

The heating flux $F_H$ into active region loops has been statistically
determined from hydrostatic simulations of the entire Sun's corona 
(Schrijver et al.~(2004) and the following scaling law was found
between the magnetic field strengths $B$ and the loop length $L$,
$$
	F_H \propto B^{1.0\pm0.3} L^{-1.0\pm0.5} \approx B L^{-1} \ .
	\eqno(B1)
$$
The heating flux $F_H$ represents the energy flux 
(in units or erg cm$^{-2}$ s$^{-1}$) per footpoint area of a loop.
From the heat flux $F_H$ we can deduce the following scaling law
for the average volumetric heating rate $H = F_H/L$,
$$
	H \propto {F_H \over L} \propto B L^{-2} \ .
	\eqno(B2)
$$
Substituting this scaling law relationship $H(B, L)$ into the
generalized size distribution $N(x)$ (Eq.~A9) with $x=B$, we can
derive the size distribution of magnetic fields for loops or flares
in the solar corona,
$$
	N(B) \propto \int N(L) N(H[B,L]) dL
	\propto \int L^{-3} (B L^{-2})^{-\alpha_H} dL
	\propto B^{-\alpha_H} 
	\left[ L^{-2+2 \alpha_H} \right]_{L,min}^{L_{max}} \ ,
	\eqno(B3)
$$
which essentially yields a size distribution $N(B) \propto B^{-1.8}$
that is identical to the heating rate distribution $N(H) \propto H^{-1.8}$,
if we neglect truncation effects (as discussed in Appendix A).

Besides this prediction for the size distributions of magnetic fields,
we can also predict the size distribution for magnetic fluxes $\Phi$ of
active regions or flaring regions with size $A \propto L^2$, which
are defined by the following scaling law,
$$
	\Phi = \int_A B(x,y) dx dy = B \ A \propto B \ L^2 \ .
	\eqno(B4)
$$
Substituting this variable of the magnetic flux $\Phi$ into the 
volumetric heating rate scaling law $H \propto B L^{-2}$ (Eq.~B2),
we have a scaling law of the heating rate $H$ as a function of
the variables $\Phi$ and $L$,
$$
	H \propto B L^{-2} \propto \Phi L^{-4} \ ,
	\eqno(B5)
$$
and can then derive the size distribution $N(\Phi)$ of magnetic fluxes
by substituting this scaling law $H(\Phi, L)$ into the
generalized size distribution $N(x)$ (Eq.~A9) with $x=\Phi$, 
which yields,
$$
	N(\Phi) \propto \int N(L) N(H[\Phi, L]) dL
	\propto \int L^{-3} (\Phi L^{-4})^{-\alpha_H} dL
	\propto \Phi^{-\alpha_H} 
	\left[ L^{-2+2 \alpha_H} \right]_{L,min}^{L_{max}} \ ,
	\eqno(B6)
$$
which essentially yields a size distribution $N(\Phi) \propto \Phi^{-1.8}$
that is identical to the magnetic field distribution $N(B) \propto B^{-1.8}$
or the heating rate distribution $N(H) \propto H^{-1.8}$,
if we neglect truncation effects (as discussed in Appendix A).
This prediction actually agrees with recent statistical observations
of the magnetic flux on the solar surface, which was found to be 
distributed as a powerlaw distribution over more than five decades 
in flux (Parnell et al.~2009),
$$
	N(\Phi ) \propto \Phi^{-1.85\pm0.14} \ ,
	\eqno(B7)
$$
based on magnetic features distributed all over the Sun, using SohO/MDI
and Hinode/SOT data. A coupling between the size distribution of
photospheric magnetic features and coronal energy dissipation events
was also established in a recent study (Uritsky et al.~2013).
Thus our result of the inferred volumetric heating
rate size distribution $N(H) \propto H^{-1.8}$ is consistent with the
statistics of magnetic fluxes on the solar surface measured by
Parnell et al.~(2009), as well as with the heating flux scaling law
$F_H = H/L \propto B L^{-1}$ found from hydrostatic simulations of the 
entire Sun's corona by Schrijver et al.~(2004).


\begin{deluxetable}{rrrrrrr}
\tabletypesize{\normalsize}
\tabletypesize{\footnotesize}
\tablecaption{Statistics of powerlaw slopes $\alpha_F$ 
of AIA fluxes $F_{\lambda}$ 
155 solar flares, tabulated in 7 AIA wavelengths and for 5 different
flux thresholds. The theoretical prediction
of the FD-SOC model is $\alpha_F=2.0$.}
\tablewidth{0pt}
\tablehead{
\colhead{Wavelength}&
\colhead{Threshold}&
\colhead{Threshold}&
\colhead{Threshold}&
\colhead{Threshold}&
\colhead{Threshold}&
\colhead{All}\\
\colhead{[$\ang$]}&
\colhead{1\%}&
\colhead{2\%}&
\colhead{5\%}&
\colhead{10\%}&
\colhead{20\%}&
\colhead{}}
\startdata
 94 &    2.2  &    2.2  &    2.2 &    2.2 &    2.1 &    2.2$\pm$   0.04\\
131 &    2.0  &    2.0  &    2.0 &    2.0 &    2.0 &    2.0$\pm$   0.02\\
171 &    2.1  &    1.9  &    2.0 &    2.0 &    2.2 &    2.0$\pm$   0.1\\
193 &    2.1  &    2.0  &    2.1 &    2.1 &    1.9 &    2.0$\pm$   0.1\\
211 &    2.3  &    2.1  &    2.1 &    2.2 &    2.0 &    2.1$\pm$   0.1\\
304 &    2.1  &    2.1  &    1.9 &    2.5 &    2.0 &    2.1$\pm$   0.2\\
335 &    1.8  &    2.1  &    2.1 &    1.8 &    1.9 &    1.9$\pm$   0.1\\
         &       &       &       &       &       &                \\
All &   2.1$\pm$  0.2 &   2.1$\pm$  0.1 &   2.1$\pm$  0.1 &   2.1$\pm$  0.2 &   2.0$\pm$  0.1 &   2.1$\pm$  0.1
\enddata
\end{deluxetable}

\begin{deluxetable}{rrrrrrrrrrr}
 \tabletypesize{\normalsize}
\tabletypesize{\footnotesize}
\tablecaption{Catalog of 155 analyzed M and X-class flare events and best-fit
model parameters: length scale $L_p$(Mm), electron temperature $T_p$ (MK), Gaussian
temperature width of DEM peak $\sigma_p$, electron density $n_p$ (cm$^{-3}$), 
emission measure peak $EM_p$ (cm$^{-3}$), thermal energy $E_{th}$ (erg), and
ratio of DEM-fitted to observed flux $F_{fit}/F_{obs}$ at the peak time of the flares.}
\tablewidth{0pt}
\tablehead{
\colhead{Nr}&
\colhead{Observation}&
\colhead{Peak}&
\colhead{GOES}&
\colhead{Length}&
\colhead{Temperature}&
\colhead{Temperature}&
\colhead{Electron}&
\colhead{Emission}&
\colhead{Thermal}&
\colhead{DEM}\\
\colhead{  }&
\colhead{date}&
\colhead{time}&
\colhead{class}&
\colhead{scale}&
\colhead{peak}&
\colhead{width}&
\colhead{density}&
\colhead{measure}&
\colhead{energy}&
\colhead{fit}\\
\colhead{  }&
\colhead{  }&
\colhead{HH:MM}&
\colhead{  }&
\colhead{$L_p$ (Mm)}&
\colhead{$T_p$ (MK)}&
\colhead{$log(\sigma_p)$}&
\colhead{$\log(n_e)$}&
\colhead{$\log(EM_p)$}&
\colhead{$\log(E_{th})$}&
\colhead{$F_{fit}/F_{obs}$}}
\startdata
   1 &2010-06-12 &00:58 &M2.0 &  21 & 14.1 &  0.50 & 10.6 &  49.3 &  30.7 &  0.96$\pm$0.26 \\
   2 &2010-06-13 &05:39 &M1.0 &  16 & 12.6 &  0.59 & 10.9 &  49.3 &  30.6 &  1.18$\pm$0.46 \\
   3 &2010-08-07 &18:24 &M1.0 &  38 &  7.9 &  0.65 & 10.1 &  48.9 &  30.7 &  1.57$\pm$0.94 \\
   4 &2010-10-16 &19:12 &M2.9 &  25 & 15.8 &  0.59 & 10.8 &  49.9 &  31.2 &  1.01$\pm$0.14 \\
   5 &2010-11-04 &23:58 &M1.6 &  15 & 10.0 &  0.60 & 10.5 &  48.5 &  30.0 &  1.37$\pm$0.83 \\
   6 &2010-11-05 &13:29 &M1.0 &  19 & 11.2 &  0.36 & 10.7 &  49.2 &  30.6 &  1.27$\pm$0.43 \\
   7 &2010-11-06 &15:36 &M5.4 &  17 & 14.1 &  0.34 &  9.6 &  47.2 &  29.4 &  0.86$\pm$0.28 \\
   8 &2011-01-28 &01:03 &M1.3 &  24 & 11.2 &  0.53 & 10.6 &  49.0 &  30.8 &  1.08$\pm$0.38 \\
   9 &2011-02-09 &01:31 &M1.9 &  17 &  7.1 &  0.66 & 10.6 &  49.1 &  30.1 &  1.25$\pm$0.58 \\
  10 &2011-02-13 &17:38 &M6.6 &  35 & 14.1 &  0.47 & 10.7 &  49.9 &  31.4 &  1.02$\pm$0.07 \\
  11 &2011-02-14 &17:26 &M2.2 &  15 &  2.5 &  0.71 &  9.9 &  48.0 &  28.8 &  1.10$\pm$0.83 \\
  12 &2011-02-15 &01:56 &X2.2 &  16 & 17.8 &  0.50 & 11.5 &  50.4 &  31.3 &  1.07$\pm$0.14 \\
  13 &2011-02-16 &01:39 &M1.0 &  12 & 15.8 &  0.37 & 11.1 &  49.1 &  30.5 &  1.13$\pm$0.37 \\
  14 &2011-02-16 &07:44 &M1.1 &  11 & 14.1 &  0.52 & 11.0 &  49.1 &  30.3 &  0.90$\pm$0.24 \\
  15 &2011-02-16 &14:25 &M1.6 &  13 & 12.6 &  0.70 & 11.3 &  49.7 &  30.7 &  1.03$\pm$0.14 \\
  16 &2011-02-18 &10:11 &M6.6 &  20 &  8.9 &  0.71 & 10.5 &  49.1 &  30.3 &  0.81$\pm$0.46 \\
  17 &2011-02-18 &10:26 &M1.0 &  29 &  7.9 &  0.69 & 10.4 &  49.0 &  30.6 &  1.04$\pm$0.42 \\
  18 &2011-02-18 &13:03 &M1.4 &  16 &  4.5 &  0.67 & 10.3 &  47.9 &  29.5 &  0.88$\pm$0.70 \\
  19 &2011-02-18 &14:08 &M1.0 &  18 &  8.9 &  0.41 & 10.4 &  48.7 &  30.1 &  1.49$\pm$0.98 \\
  20 &2011-02-18 &21:04 &M1.3 &   7 & 14.1 &  0.55 & 10.9 &  49.2 &  29.7 &  1.11$\pm$0.36 \\
  21 &2011-02-24 &07:35 &M3.5 &   8 & 14.1 &  0.22 & 11.3 &  49.4 &  30.2 &  1.44$\pm$0.86 \\
  22 &2011-02-28 &12:52 &M1.1 &  14 & 12.6 &  0.50 & 11.0 &  49.3 &  30.5 &  0.98$\pm$0.22 \\
  23 &2011-03-07 &05:13 &M1.2 &  21 & 12.6 &  0.56 & 10.7 &  49.3 &  30.7 &  1.01$\pm$0.22 \\
  24 &2011-03-07 &07:54 &M1.5 &   4 & 14.1 &  0.51 & 11.4 &  49.1 &  29.4 &  1.14$\pm$0.44 \\
  25 &2011-03-07 &08:07 &M1.4 &  36 & 12.6 &  0.49 & 10.5 &  49.2 &  31.2 &  1.04$\pm$0.48 \\
  26 &2011-03-07 &09:20 &M1.8 &  11 & 11.2 &  0.22 & 10.2 &  47.0 &  29.3 &  1.03$\pm$0.93 \\
  27 &2011-03-07 &14:30 &M1.9 &  25 &  8.9 &  0.40 & 10.7 &  49.4 &  30.8 &  1.19$\pm$0.44 \\
  28 &2011-03-07 &20:12 &M3.7 &  63 & 10.0 &  0.24 & 10.4 &  49.6 &  31.7 &  1.27$\pm$0.44 \\
  29 &2011-03-07 &21:50 &M1.5 &  12 & 10.0 &  0.71 & 11.0 &  49.2 &  30.2 &  1.06$\pm$0.13 \\
  30 &2011-03-08 &02:29 &M1.3 &   9 &  6.3 &  0.73 & 11.1 &  49.0 &  29.8 &  1.00$\pm$0.14 \\
  31 &2011-03-08 &03:58 &M1.5 &  16 &  8.9 &  0.64 & 10.6 &  49.0 &  30.2 &  1.30$\pm$0.80 \\
  32 &2011-03-08 &10:44 &M5.3 &  28 & 17.8 &  0.55 & 10.8 &  49.8 &  31.3 &  1.01$\pm$0.13 \\
  33 &2011-03-08 &18:28 &M4.4 &  10 & 14.1 &  0.25 & 11.2 &  49.3 &  30.4 &  0.91$\pm$0.20 \\
  34 &2011-03-08 &20:16 &M1.4 &  12 & 10.0 &  0.65 & 10.3 &  48.4 &  29.6 &  0.69$\pm$0.42 \\
  35 &2011-03-09 &11:07 &M1.7 &  32 &  5.6 &  0.69 & 10.3 &  48.0 &  30.5 &  1.07$\pm$0.87 \\
  36 &2011-03-09 &14:02 &M1.7 &  18 & 10.0 &  0.55 & 10.4 &  48.3 &  30.2 &  1.01$\pm$0.48 \\
  37 &2011-03-09 &23:23 &X1.5 &  16 & 17.8 &  0.46 & 11.5 &  50.4 &  31.3 &  1.09$\pm$0.26 \\
  38 &2011-03-10 &22:41 &M1.1 &   8 &  8.9 &  0.72 & 10.6 &  47.9 &  29.3 &  1.03$\pm$0.50 \\
  39 &2011-03-12 &04:43 &M1.3 &  14 & 11.2 &  0.53 & 11.2 &  49.4 &  30.7 &  0.97$\pm$0.25 \\
  40 &2011-03-14 &19:52 &M4.2 &   8 & 17.8 &  0.54 & 11.5 &  49.8 &  30.4 &  1.03$\pm$0.06 \\
  41 &2011-03-15 &00:22 &M1.0 &   6 & 10.0 &  0.73 & 11.2 &  49.0 &  29.6 &  1.00$\pm$0.25 \\
  42 &2011-03-23 &02:17 &M1.4 &  12 & 14.1 &  0.49 & 11.1 &  49.4 &  30.5 &  1.05$\pm$0.11 \\
  43 &2011-03-24 &12:07 &M1.0 &  12 & 10.0 &  0.62 & 11.2 &  49.3 &  30.4 &  0.99$\pm$0.09 \\
  44 &2011-03-25 &23:22 &M1.0 &  14 & 11.2 &  0.36 & 10.8 &  49.2 &  30.4 &  1.12$\pm$0.31 \\
  45 &2011-04-15 &17:12 &M1.3 &   7 & 11.2 &  0.55 & 11.1 &  48.9 &  29.7 &  0.93$\pm$0.40 \\
  46 &2011-04-22 &04:57 &M1.8 &  14 & 11.2 &  0.33 & 11.0 &  49.3 &  30.5 &  1.14$\pm$0.40 \\
  47 &2011-04-22 &15:53 &M1.2 &  11 & 15.8 &  0.60 & 11.3 &  49.4 &  30.6 &  1.02$\pm$0.19 \\
  48 &2011-05-28 &21:50 &M1.1 &  24 & 12.6 &  0.52 & 10.7 &  49.2 &  30.9 &  1.14$\pm$0.31 \\
  49 &2011-05-29 &10:33 &M1.4 &  16 &  8.9 &  0.64 & 10.7 &  48.8 &  30.3 &  1.30$\pm$0.88 \\
  50 &2011-06-07 &06:41 &M2.5 &  52 &  7.1 &  0.63 & 10.2 &  49.2 &  31.2 &  0.70$\pm$0.49 \\
  51 &2011-06-14 &21:47 &M1.3 &  25 & 12.6 &  0.50 & 10.6 &  49.3 &  30.8 &  0.97$\pm$0.26 \\
  52 &2011-07-27 &16:07 &M1.1 &  24 & 11.2 &  0.25 & 10.4 &  48.9 &  30.6 &  1.18$\pm$0.48 \\
  53 &2011-07-30 &02:09 &M9.3 &  35 & 15.8 &  0.55 & 10.7 &  50.2 &  31.5 &  1.00$\pm$0.10 \\
  54 &2011-08-02 &06:19 &M1.4 &  34 & 10.0 &  0.37 & 10.4 &  49.3 &  30.9 &  1.15$\pm$0.40 \\
  55 &2011-08-03 &03:37 &M1.1 &  25 & 12.6 &  0.40 & 10.6 &  49.2 &  30.8 &  1.17$\pm$0.35 \\
  56 &2011-08-03 &04:32 &M1.7 &  31 & 12.6 &  0.70 & 10.6 &  49.6 &  31.1 &  0.98$\pm$0.10 \\
  57 &2011-08-03 &13:48 &M6.0 &  37 & 12.6 &  0.42 & 10.6 &  49.9 &  31.3 &  1.09$\pm$0.36 \\
  58 &2011-08-04 &03:57 &M9.3 &  45 & 12.6 &  0.46 & 10.6 &  50.1 &  31.7 &  1.09$\pm$0.22 \\
  59 &2011-08-08 &18:10 &M3.5 &  28 & 11.2 &  0.53 & 10.6 &  49.4 &  30.9 &  0.95$\pm$0.35 \\
  60 &2011-08-09 &03:54 &M2.5 &  24 & 14.1 &  0.46 & 10.7 &  49.4 &  30.9 &  1.02$\pm$0.17 \\
  61 &2011-08-09 &08:05 &X6.9 &   9 & 17.8 &  0.46 & 11.9 &  50.6 &  31.1 &  1.05$\pm$0.23 \\
  62 &2011-09-04 &11:45 &M3.2 &  21 & 15.8 &  0.36 & 10.8 &  49.3 &  30.9 &  1.42$\pm$0.70 \\
  63 &2011-09-05 &04:28 &M1.6 &  22 &  8.9 &  0.38 & 10.3 &  48.7 &  30.2 &  1.28$\pm$0.79 \\
  64 &2011-09-05 &07:58 &M1.2 &  18 &  8.9 &  0.59 & 10.4 &  48.5 &  30.1 &  1.20$\pm$0.81 \\
  65 &2011-09-06 &01:50 &M5.3 &  29 & 14.1 &  0.45 & 10.7 &  49.8 &  31.2 &  1.06$\pm$0.17 \\
  66 &2011-09-06 &22:20 &X2.1 &  50 & 14.1 &  0.59 & 10.6 &  50.2 &  31.8 &  1.05$\pm$0.32 \\
  67 &2011-09-07 &22:38 &X1.8 &  57 & 15.8 &  0.56 & 10.6 &  50.3 &  32.0 &  0.99$\pm$0.08 \\
  68 &2011-09-08 &15:46 &M6.7 &  42 & 15.8 &  0.57 & 10.6 &  50.0 &  31.7 &  1.05$\pm$0.22 \\
  69 &2011-09-09 &06:11 &M2.7 &  22 & 14.1 &  0.41 & 10.8 &  49.6 &  31.0 &  1.37$\pm$0.73 \\
  70 &2011-09-09 &12:49 &M1.2 &  26 & 12.6 &  0.51 & 10.7 &  49.3 &  31.0 &  1.18$\pm$0.47 \\
  71 &2011-09-10 &07:40 &M1.1 &  23 & 11.2 &  0.62 & 10.6 &  49.0 &  30.7 &  0.90$\pm$0.47 \\
  72 &2011-09-21 &12:23 &M1.8 &  23 & 12.6 &  0.65 & 10.4 &  48.7 &  30.5 &  1.25$\pm$0.66 \\
  73 &2011-09-22 &10:00 &M1.1 &  10 & 14.1 &  0.49 & 11.1 &  48.9 &  30.3 &  1.11$\pm$0.62 \\
  74 &2011-09-22 &11:01 &X1.4 &  22 & 17.8 &  0.39 & 11.1 &  50.2 &  31.4 &  0.96$\pm$0.22 \\
  75 &2011-09-23 &01:59 &M1.6 &  22 & 14.1 &  0.41 & 10.6 &  49.1 &  30.7 &  1.24$\pm$0.50 \\
  76 &2011-09-23 &22:15 &M1.6 &   9 & 11.2 &  0.29 & 11.0 &  49.0 &  29.9 &  1.10$\pm$0.29 \\
  77 &2011-09-23 &23:56 &M1.9 &  32 & 11.2 &  0.55 & 10.6 &  49.3 &  31.1 &  1.08$\pm$0.22 \\
  78 &2011-09-24 &09:40 &X1.9 &  10 & 17.8 &  0.45 & 11.9 &  50.0 &  31.1 &  1.04$\pm$0.34 \\
  79 &2011-09-24 &13:20 &M7.1 &  27 & 11.2 &  0.30 & 10.9 &  49.7 &  31.2 &  1.06$\pm$0.21 \\
  80 &2011-09-24 &16:59 &M1.7 &  19 & 12.6 &  0.22 & 10.5 &  48.7 &  30.4 &  1.09$\pm$0.39 \\
  81 &2011-09-24 &17:25 &M3.1 &   7 &  7.1 &  0.68 & 10.9 &  48.4 &  29.3 &  1.45$\pm$0.65 \\
  82 &2011-09-24 &18:15 &M2.8 &  14 & 11.2 &  0.45 & 11.0 &  49.0 &  30.5 &  1.12$\pm$0.35 \\
  83 &2011-09-24 &19:21 &M3.0 &  17 & 12.6 &  0.34 & 11.0 &  49.4 &  30.8 &  1.09$\pm$0.27 \\
  84 &2011-09-24 &20:36 &M5.8 &   6 & 15.8 &  0.47 & 11.4 &  49.5 &  29.9 &  1.34$\pm$0.75 \\
  85 &2011-09-24 &21:27 &M1.2 &  10 &  6.3 &  0.62 & 10.8 &  48.5 &  29.6 &  1.22$\pm$0.86 \\
  86 &2011-09-24 &23:58 &M1.0 &  15 &  8.9 &  0.68 & 10.3 &  47.8 &  29.8 &  0.66$\pm$0.49 \\
  87 &2011-09-25 &02:33 &M4.4 &   5 & 14.1 &  0.36 & 11.1 &  48.6 &  29.5 &  1.23$\pm$0.51 \\
  88 &2011-09-25 &04:50 &M7.4 &  19 & 12.6 &  0.36 & 11.0 &  49.8 &  30.9 &  1.10$\pm$0.18 \\
  89 &2011-09-25 &08:49 &M3.1 &   5 & 15.8 &  0.54 & 11.4 &  49.3 &  29.8 &  1.11$\pm$0.27 \\
  90 &2011-09-25 &09:35 &M1.5 &   9 &  1.3 &  0.23 & 10.3 &  47.0 &  28.2 &  4.82$\pm$8.82 \\
  91 &2011-09-25 &15:33 &M3.7 &  16 & 11.2 &  0.71 & 11.2 &  49.8 &  30.8 &  1.03$\pm$0.15 \\
  92 &2011-09-25 &16:58 &M2.2 &   7 & 14.1 &  0.44 & 11.2 &  49.3 &  30.0 &  1.07$\pm$0.14 \\
  93 &2011-09-26 &05:08 &M4.0 &  10 & 15.8 &  0.54 & 11.5 &  49.8 &  30.8 &  0.98$\pm$0.12 \\
  94 &2011-09-26 &14:46 &M2.6 &  21 & 12.6 &  0.69 & 11.2 &  49.9 &  31.2 &  1.07$\pm$0.13 \\
  95 &2011-09-28 &13:28 &M1.2 &  12 & 12.6 &  0.74 & 11.2 &  49.6 &  30.5 &  1.04$\pm$0.16 \\
  96 &2011-09-30 &19:06 &M1.0 &  11 & 10.0 &  0.32 & 10.9 &  49.0 &  30.0 &  1.13$\pm$0.26 \\
  97 &2011-10-01 &09:59 &M1.2 &  40 &  8.9 &  0.39 & 10.5 &  49.3 &  31.2 &  1.22$\pm$0.41 \\
  98 &2011-10-02 &00:50 &M3.9 &  13 & 15.8 &  0.41 & 11.5 &  49.7 &  31.0 &  1.02$\pm$0.19 \\
  99 &2011-10-02 &17:23 &M1.3 &   8 &  7.9 &  0.75 & 11.2 &  49.2 &  29.9 &  0.98$\pm$0.16 \\
 100 &2011-10-20 &03:25 &M1.6 &  21 &  2.8 &  0.68 & 10.4 &  47.9 &  29.8 &  1.33$\pm$0.81 \\
 101 &2011-10-21 &13:00 &M1.3 &  11 &  8.9 &  0.66 & 11.1 &  49.1 &  30.1 &  1.20$\pm$0.79 \\
 102 &2011-10-22 &11:10 &M1.3 &  35 &  6.3 &  0.64 &  9.8 &  48.6 &  30.2 &  1.06$\pm$0.80 \\
 103 &2011-10-31 &15:08 &M1.1 &   6 &  8.9 &  0.64 & 10.5 &  47.7 &  28.7 &  0.85$\pm$0.71 \\
 104 &2011-10-31 &18:08 &M1.4 &  14 & 11.2 &  0.22 & 10.7 &  48.7 &  30.2 &  0.89$\pm$0.38 \\
 105 &2011-11-02 &22:01 &M4.3 &   6 & 15.8 &  0.45 & 11.4 &  49.6 &  30.0 &  1.35$\pm$0.78 \\
 106 &2011-11-03 &11:11 &M2.5 &   6 & 15.8 &  0.40 & 11.4 &  49.4 &  30.0 &  1.33$\pm$0.47 \\
 107 &2011-11-03 &20:27 &X1.9 &  10 & 15.8 &  0.56 & 11.7 &  50.2 &  30.9 &  1.08$\pm$0.26 \\
 108 &2011-11-03 &23:36 &M2.1 &  16 &  8.9 &  0.64 & 10.4 &  48.3 &  30.0 &  1.07$\pm$0.82 \\
 109 &2011-11-04 &20:40 &M1.0 &   8 & 15.8 &  0.39 & 11.0 &  49.0 &  30.0 &  1.35$\pm$0.59 \\
 110 &2011-11-05 &03:35 &M3.7 &  25 & 14.1 &  0.38 & 10.6 &  49.6 &  30.9 &  1.11$\pm$0.27 \\
 111 &2011-11-05 &11:21 &M1.1 &  15 & 14.1 &  0.50 & 10.6 &  48.8 &  30.3 &  1.31$\pm$0.92 \\
 112 &2011-11-05 &20:38 &M1.8 &  13 & 12.6 &  0.42 & 11.1 &  49.2 &  30.5 &  0.99$\pm$0.16 \\
 113 &2011-11-06 &01:03 &M1.2 &  10 & 11.2 &  0.71 & 11.0 &  49.2 &  30.1 &  0.99$\pm$0.13 \\
 114 &2011-11-06 &06:35 &M1.4 &  14 & 12.6 &  0.37 & 11.0 &  49.2 &  30.5 &  1.12$\pm$0.23 \\
 115 &2011-11-09 &13:35 &M1.1 &  50 &  7.9 &  0.31 & 10.2 &  49.1 &  31.2 &  1.26$\pm$0.75 \\
 116 &2011-11-15 &09:12 &M1.2 &   8 & 11.2 &  0.46 & 11.2 &  49.0 &  29.9 &  1.01$\pm$0.14 \\
 117 &2011-11-15 &12:43 &M1.9 &   9 & 12.6 &  0.45 & 11.3 &  49.5 &  30.2 &  1.11$\pm$0.27 \\
 118 &2011-11-15 &22:35 &M1.1 &  18 & 10.0 &  0.49 & 10.7 &  48.6 &  30.4 &  1.05$\pm$0.41 \\
 119 &2011-12-25 &18:16 &M4.0 &  26 & 12.6 &  0.65 & 10.7 &  49.7 &  31.0 &  1.03$\pm$0.34 \\
 120 &2011-12-26 &02:27 &M1.5 &  17 & 11.2 &  0.26 & 10.7 &  49.0 &  30.4 &  1.05$\pm$0.27 \\
 121 &2011-12-26 &20:30 &M2.3 &  21 & 14.1 &  0.46 & 10.8 &  49.5 &  30.9 &  0.97$\pm$0.23 \\
 122 &2011-12-29 &13:50 &M1.9 &  16 & 14.1 &  0.56 & 10.9 &  49.2 &  30.6 &  1.37$\pm$0.89 \\
 123 &2011-12-29 &21:51 &M2.0 &  14 & 12.6 &  0.59 & 10.9 &  49.2 &  30.3 &  1.20$\pm$0.69 \\
 124 &2011-12-30 &03:09 &M1.2 &  12 & 11.2 &  0.64 & 10.8 &  48.9 &  30.1 &  0.89$\pm$0.42 \\
 125 &2011-12-31 &13:15 &M2.4 &  17 & 10.0 &  0.58 & 10.6 &  48.9 &  30.3 &  1.08$\pm$0.56 \\
 126 &2011-12-31 &16:26 &M1.5 &  22 & 11.2 &  0.48 & 10.6 &  49.2 &  30.6 &  1.15$\pm$0.37 \\
 127 &2012-01-14 &13:18 &M1.4 &  12 &  6.3 &  0.58 & 10.5 &  48.3 &  29.5 &  1.28$\pm$0.79 \\
 128 &2012-01-17 &04:53 &M1.0 &  22 & 10.0 &  0.58 & 10.7 &  49.2 &  30.7 &  1.03$\pm$0.27 \\
 129 &2012-01-18 &19:12 &M1.7 &  31 & 14.1 &  0.54 & 10.5 &  49.4 &  31.1 &  0.95$\pm$0.30 \\
 130 &2012-01-19 &16:05 &M3.2 &  44 &  8.9 &  0.24 & 10.2 &  49.5 &  31.1 &  1.09$\pm$0.40 \\
 131 &2012-01-23 &03:59 &M8.7 &  45 & 15.8 &  0.39 & 10.5 &  50.0 &  31.6 &  1.20$\pm$0.55 \\
 132 &2012-01-27 &18:37 &X1.7 &  51 & 10.0 &  0.41 & 10.4 &  49.9 &  31.5 &  1.07$\pm$0.54 \\
 133 &2012-02-06 &20:00 &M1.0 &  31 & 11.2 &  0.56 & 10.3 &  49.4 &  30.8 &  0.97$\pm$0.36 \\
 134 &2012-03-02 &17:46 &M3.3 &  16 & 17.8 &  0.30 & 10.7 &  49.1 &  30.5 &  1.01$\pm$0.63 \\
 135 &2012-03-04 &10:52 &M2.0 &  21 & 12.6 &  0.23 & 10.7 &  49.1 &  30.7 &  1.15$\pm$0.57 \\
 136 &2012-03-05 &04:09 &X1.1 &  43 & 10.0 &  0.41 & 10.4 &  49.7 &  31.3 &  1.13$\pm$0.54 \\
 137 &2012-03-05 &19:16 &M2.1 &  16 & 11.2 &  0.59 & 10.7 &  49.0 &  30.4 &  1.02$\pm$0.45 \\
 138 &2012-03-05 &19:30 &M1.8 &  17 & 10.0 &  0.57 & 10.8 &  48.7 &  30.5 &  1.05$\pm$0.67 \\
 139 &2012-03-05 &22:34 &M1.3 &  14 &  6.3 &  0.61 & 10.0 &  47.4 &  29.2 &  1.49$\pm$0.89 \\
 140 &2012-03-06 &00:28 &M1.3 &  15 &  8.9 &  0.45 & 10.6 &  48.7 &  30.0 &  1.54$\pm$0.90 \\
 141 &2012-03-06 &01:44 &M1.2 &  13 &  8.9 &  0.62 & 10.7 &  48.7 &  30.0 &  1.03$\pm$0.48 \\
 142 &2012-03-06 &04:05 &M1.0 &  18 &  7.1 &  0.66 & 10.5 &  48.6 &  30.1 &  1.23$\pm$0.45 \\
 143 &2012-03-06 &07:55 &M1.0 &  17 &  8.9 &  0.47 & 10.7 &  48.5 &  30.3 &  1.46$\pm$0.96 \\
 144 &2012-03-06 &12:41 &M2.1 &  20 & 14.1 &  0.59 & 10.7 &  49.4 &  30.8 &  0.99$\pm$0.20 \\
 145 &2012-03-06 &21:11 &M1.3 &  14 &  8.9 &  0.48 & 10.6 &  48.8 &  30.0 &  1.32$\pm$0.86 \\
 146 &2012-03-06 &22:53 &M1.0 &  11 & 15.8 &  0.44 & 10.7 &  48.8 &  30.0 &  1.07$\pm$0.26 \\
 147 &2012-03-07 &00:24 &X5.4 &  60 & 14.1 &  0.52 & 10.6 &  50.4 &  32.0 &  1.12$\pm$0.28 \\
 148 &2012-03-07 &01:14 &X1.3 &  33 & 15.8 &  0.43 & 10.4 &  49.6 &  31.1 &  1.25$\pm$0.54 \\
 149 &2012-03-09 &03:53 &M6.3 &  38 &  7.9 &  0.28 & 10.4 &  49.5 &  31.0 &  1.25$\pm$0.47 \\
 150 &2012-03-10 &17:44 &M8.4 &  46 & 11.2 &  0.37 & 10.5 &  49.8 &  31.4 &  1.09$\pm$0.22 \\
 151 &2012-03-13 &17:41 &M7.9 &  37 & 11.2 &  0.31 & 10.6 &  49.8 &  31.3 &  1.31$\pm$0.42 \\
 152 &2012-03-14 &15:21 &M2.8 &  27 & 10.0 &  0.48 & 10.5 &  49.4 &  30.7 &  1.04$\pm$0.34 \\
 153 &2012-03-15 &07:52 &M1.8 &  28 &  5.6 &  0.63 & 10.7 &  49.7 &  30.7 &  1.16$\pm$0.37 \\
 154 &2012-03-17 &20:39 &M1.3 &  15 & 12.6 &  0.63 & 10.9 &  49.1 &  30.5 &  0.95$\pm$0.40 \\
 155 &2012-03-23 &19:40 &M1.0 &  14 & 11.2 &  0.62 & 10.7 &  49.0 &  30.2 &  1.17$\pm$0.55 \\
\enddata
\end{deluxetable}

\clearpage

\begin{figure}
\plotone{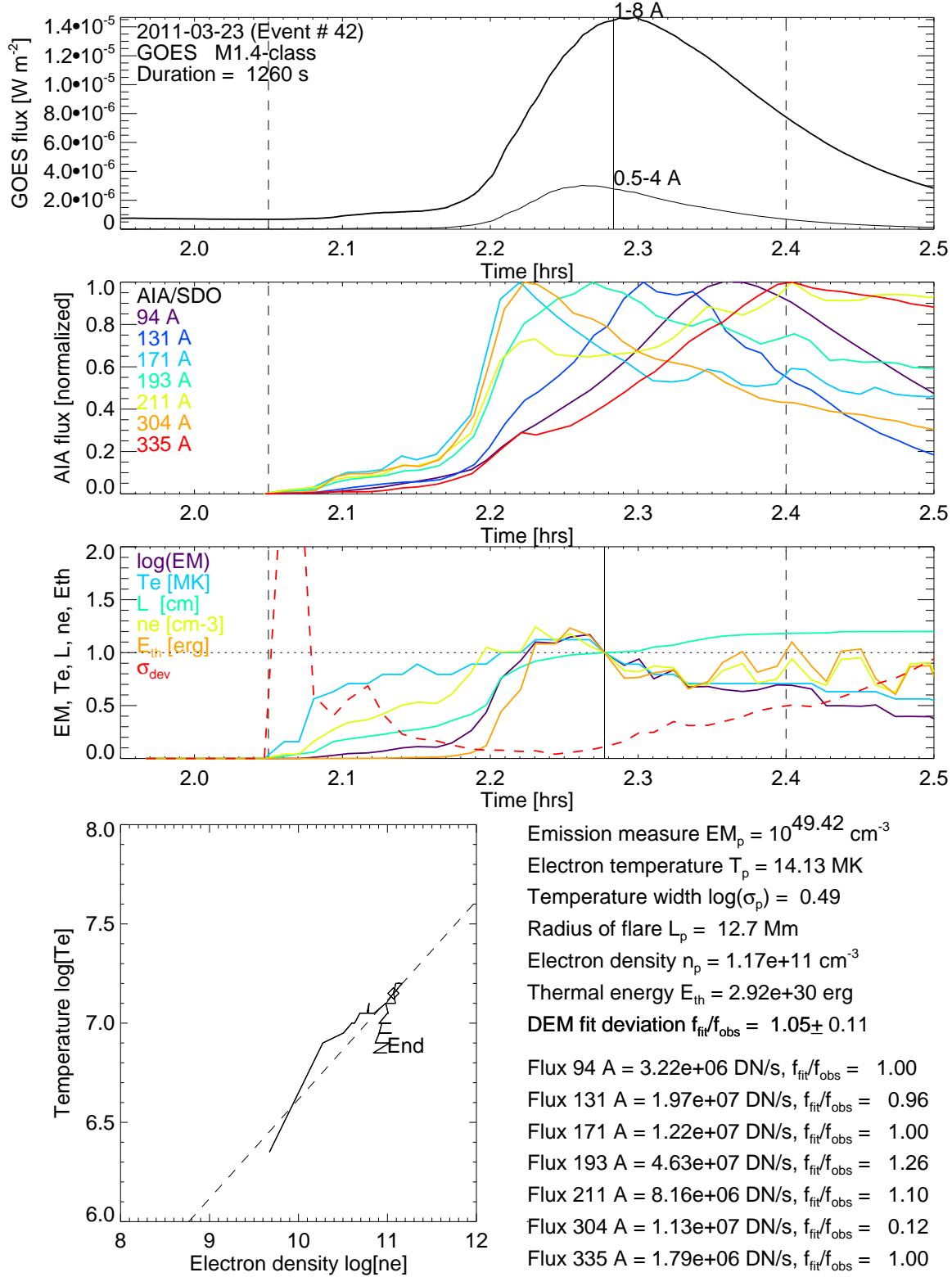}
\caption{{\sl Top panel:} GOES time profiles, with time in hrs [UT];
{\sl Second panel:} AIA time profiles;
{\sl Third panel:} Time profiles of peak emission measure $EM_p(t)$,
peak temperature $T_e(t)$, length scale $L(t)$, electron density $n_e(t)$,
thermal energy $E_{th}(t)$, and DEM fit quality $\sigma_{dev}$;
{\sl Fourth panel:} Evolutionary phase diagram $T_p(n_e)$. 
{\sl Bottom right:} Physical parameters at the flare peak time $t_p$,
observed total fluxes $f_{obs}$, and DEM fit ratios $f_{fit}/f_{obs}$,
fitted at the time of the GOES 1-8 \ang\ peak, here at 2011-03-12, 02:17 UT.}
\end{figure}

\begin{figure}
\plotone{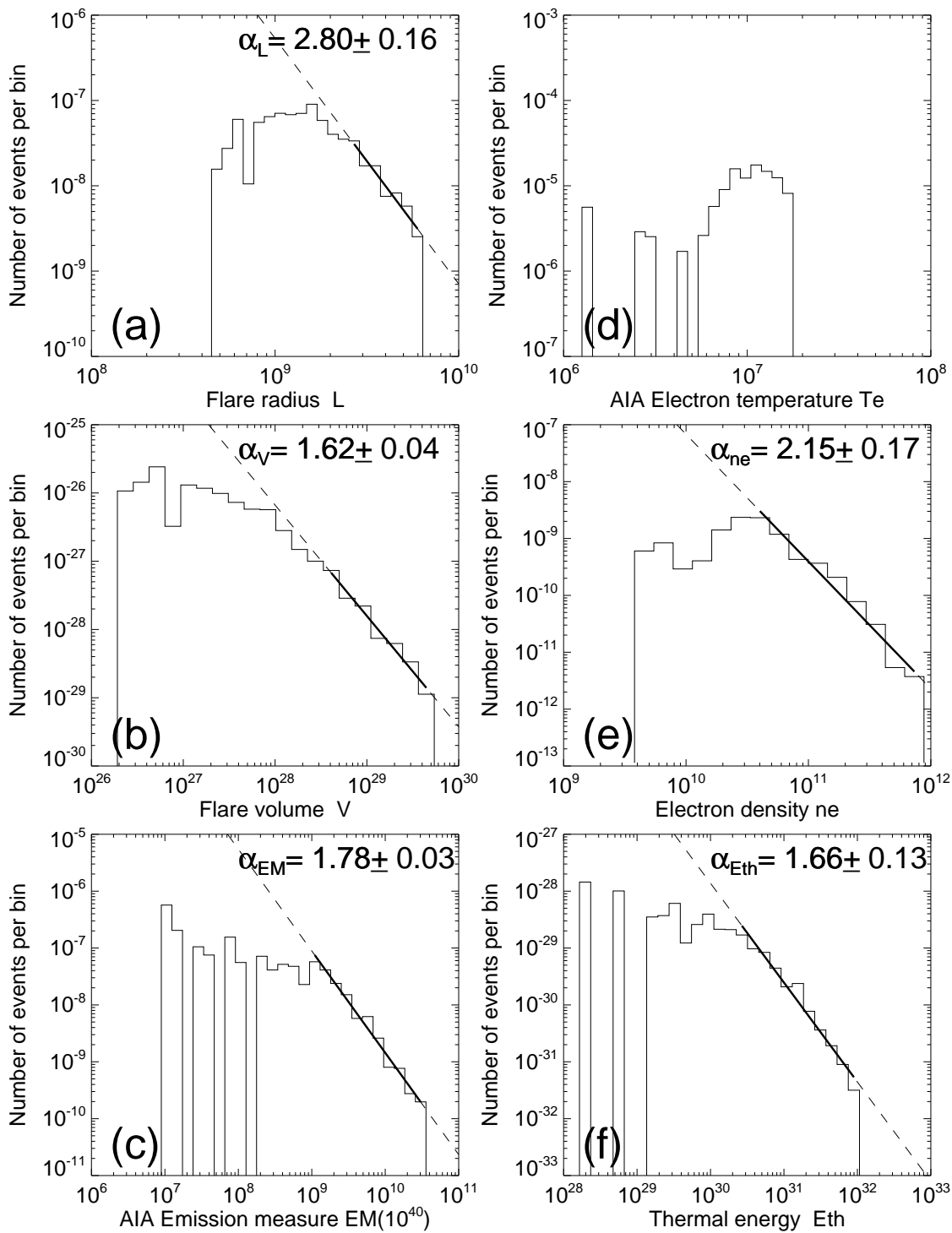}
\caption{Size distributions of the flare length scale $L_p$, flare volume $V_p$,
total emission measure $EM_p$, electron temperature $T_p$, electron
density $n_p$, and thermal energy $E_{th}$ at the flare peak times $t_p$
of the analyzed 155 M- and X-class flares observed with AIA/SDO.
Powerlaw functions are fitted at the upper end of the distributions,
and the slopes and uncertainties (inferred from 5 different bin widths used
for the powerlaw fits) are indicated.}
\end{figure}

\begin{figure}
\plotone{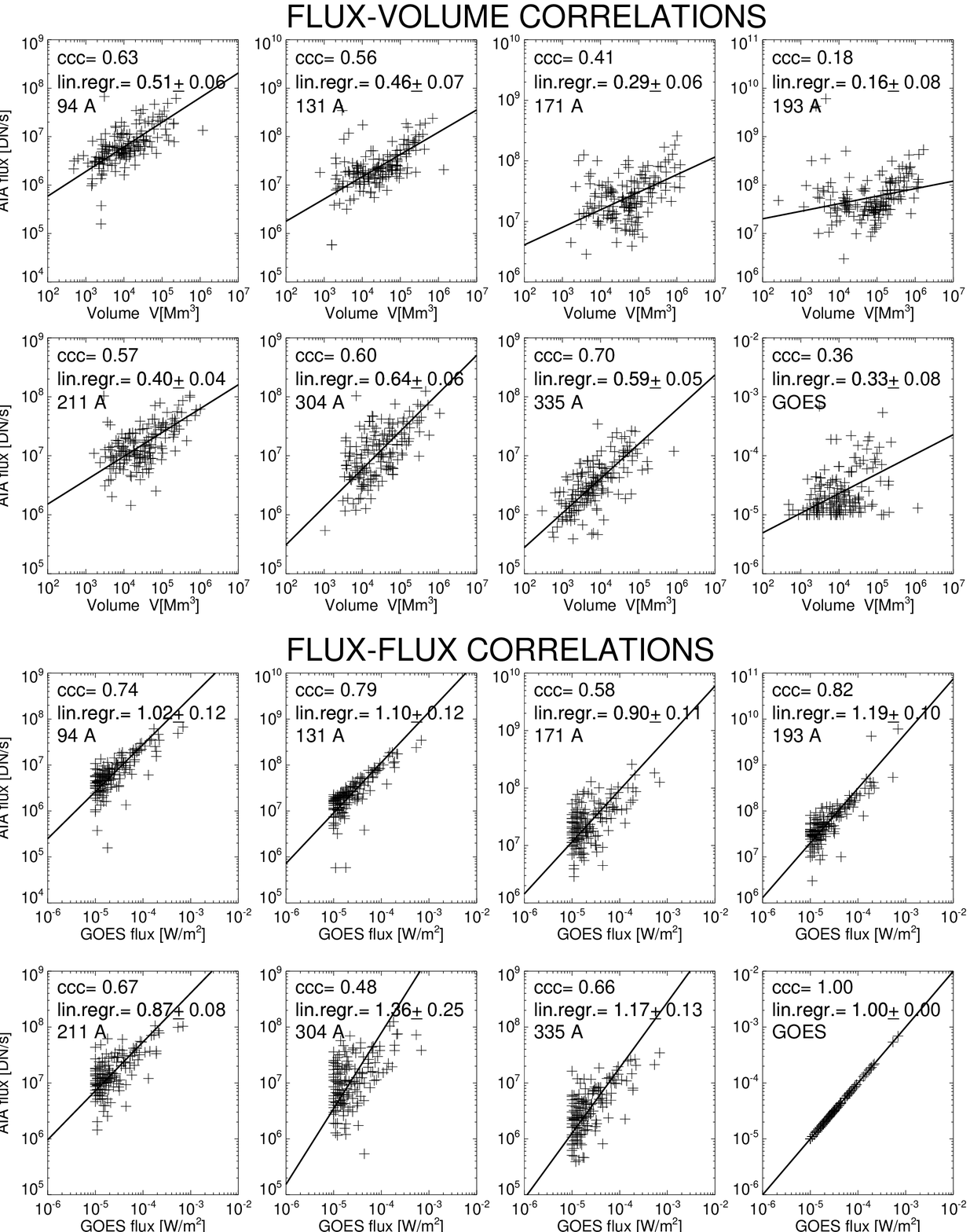}
\caption{Flux-volume correlations $F_{\lambda} \propto V^\gamma$
(first and second row) and flux-flux correlations $F_{AIA,\lambda}
\propto (F_{GOES})^\delta$ (third and bottom row) for the flare peak fluxes
$F_\lambda$ of the 155 analyzed M and X-class flares observed with
AIA/SDO and GOES. The cross-correlation coefficients (ccc) and linear
regression fits are also indicated.}
\end{figure}

\begin{figure}
\plotone{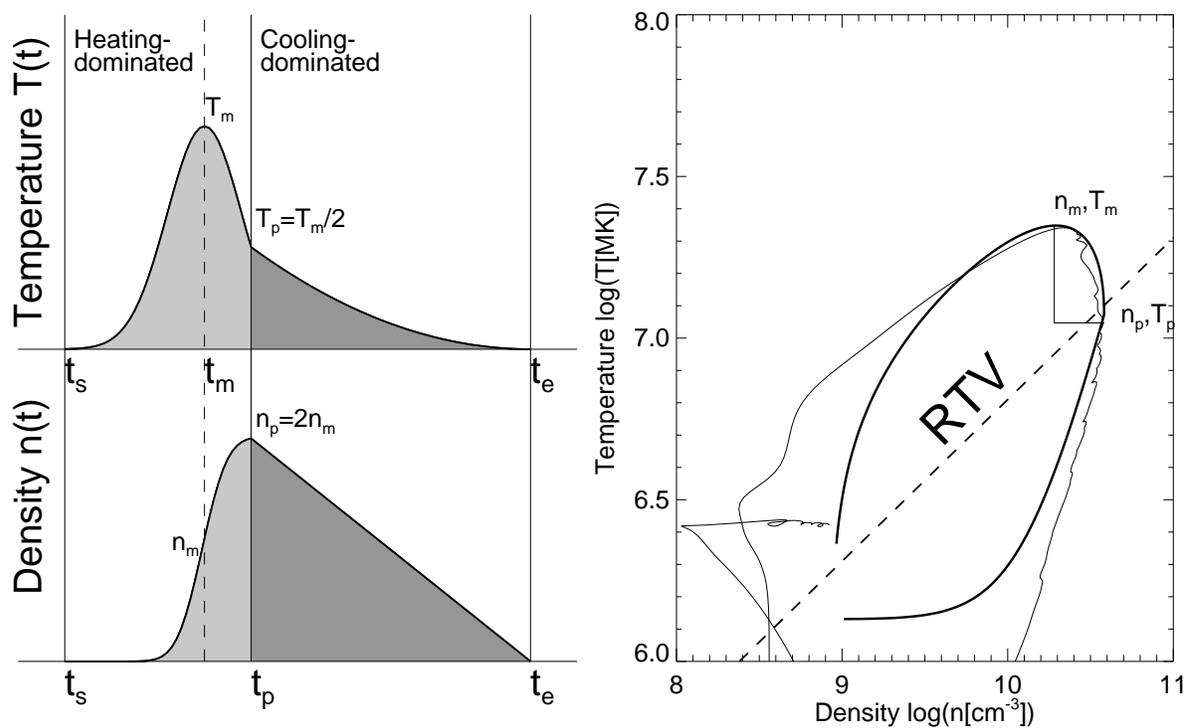}
\caption{Hydrodynamic time evolution of the electron temperature $T(t)$
and density $n_e(t)$ of a simulation of an impulsively-heated flare loop 
(see Aschwanden and Tsiklauri 2009), shown as time profiles (left panel) 
and as an evolutionary phase diagram $T_e(n_e)$ (right panel). The evolution
of the hydrodynamic simulation is shown as exact numerical solution (curve 
with thin linestyle in right panel), and as an analytical approximation
(curves with thick linestyle in both panels, along with the prediction
$T_e \propto n_e^{1/2}$ of the RTV scaling law for uniform steady 
heating (dashed line in right panel).}
\end{figure}

\begin{figure}
\plotone{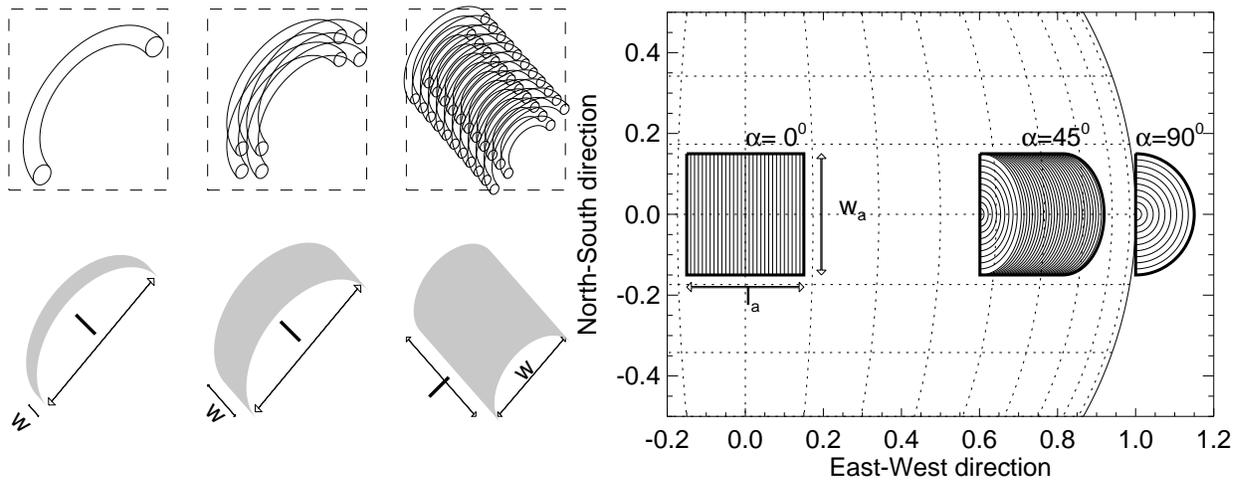}
\caption{Geometric models of multi-loop flare geometries are pictured
in terms of observed length $l$ and width $w$ parameters:
single-loop model with width $w \ll l$ (first left panel),
small multi-loop model with $w \lapprox l$ (second left panel),
large multi-loop arcade model with $w \le l$ (third left panel),
and a semi-cylindrical flare arcade at various aspect angles 
from Sun center to limb (right panel).}
\end{figure}

\begin{figure}
\plotone{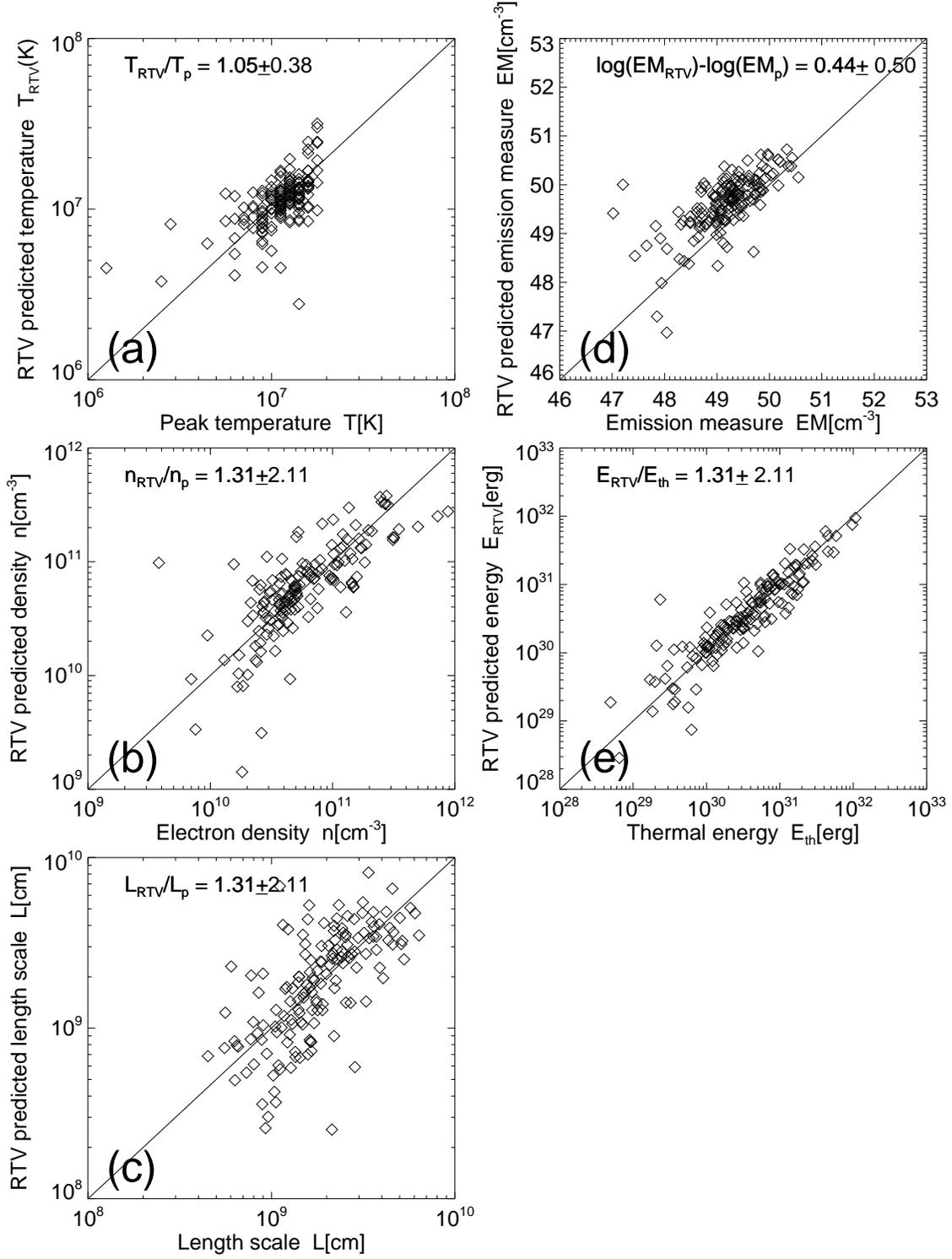}
\caption{Comparison of observed physical parameters 
$(T_p, n_p, L_p, EM_p, E_{th})$ with RTV-predicted values
according to Eqs.~(13)-(17). The mean ratio and standard deviations
of the RTV-predicted values to the observed values is also indicated.
The diagonal line represents the RTV prediction in absolute values
without adjustment parameter. Note that the ratios are close to unity.}
\end{figure}

\begin{figure}
\plotone{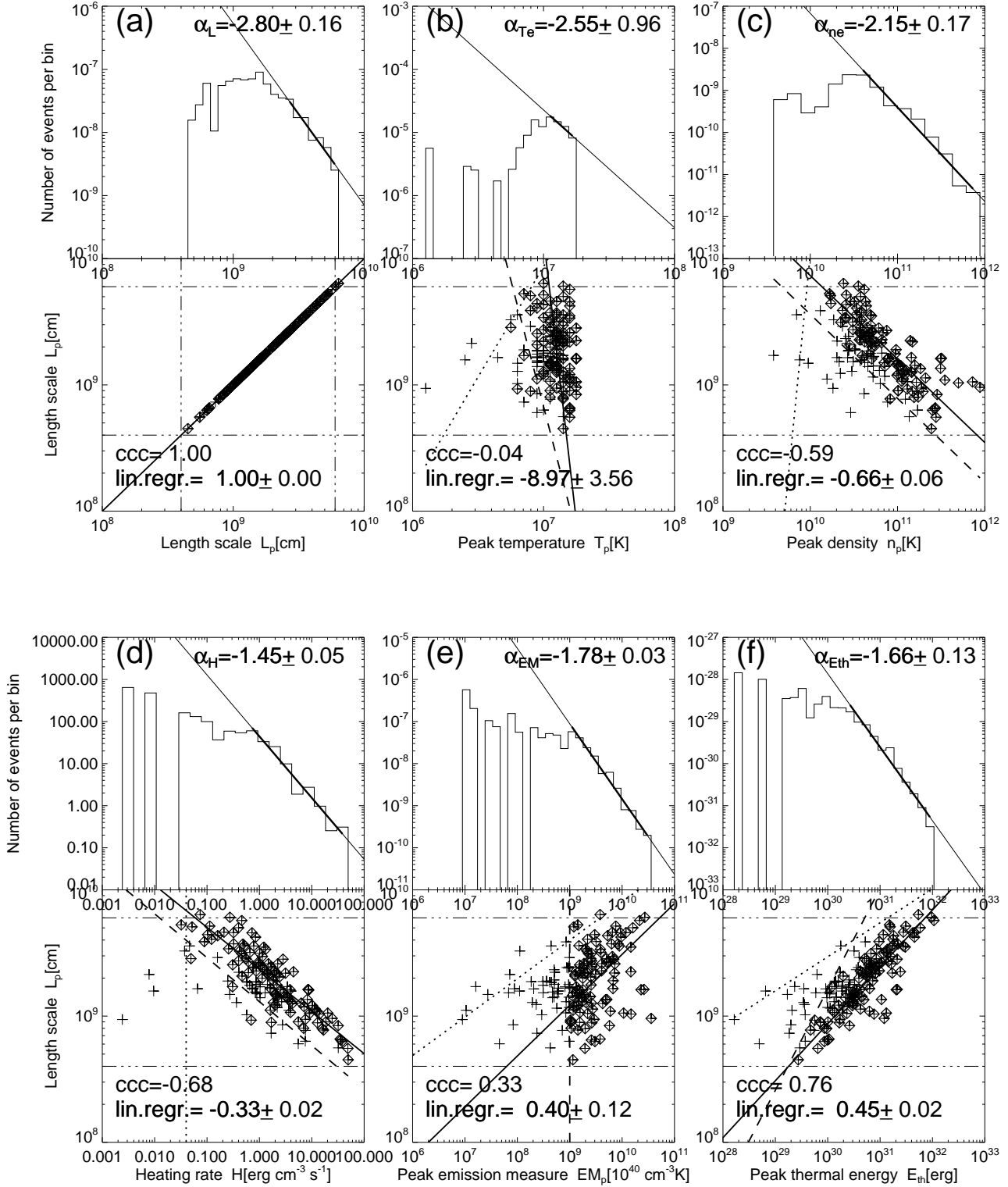}
\caption{Correlations (scatterplots) and size distributions (histograms)
for the observed peak temperatures $T_p$ (top middle), peak densities
$n_p$ (top right), heating rates $H$ (bottom left),
peak emission measures $EM_p$ (bottom middle), and
peak thermal energies $E_{th}$. The histograms are shown with powerlaw
fits (solid linestyles), and the correlations are shown with a 
linear regression fit (thick solid line), the threshold sensitivity 
is indicated for a minimum emission measure $EM_0$ (dashed linestyles) 
and for a minimum heating rate $H_0$ (dotted linestyles).
The parameters of the observed 155 flare events are shown with two 
different symbols: diamonds for the events with an emission measure 
above the threshold, $EM \ge EM_0$, and crosses for events below 
the threshold, $EM < EM_0$.}
\end{figure}

\begin{figure}
\plotone{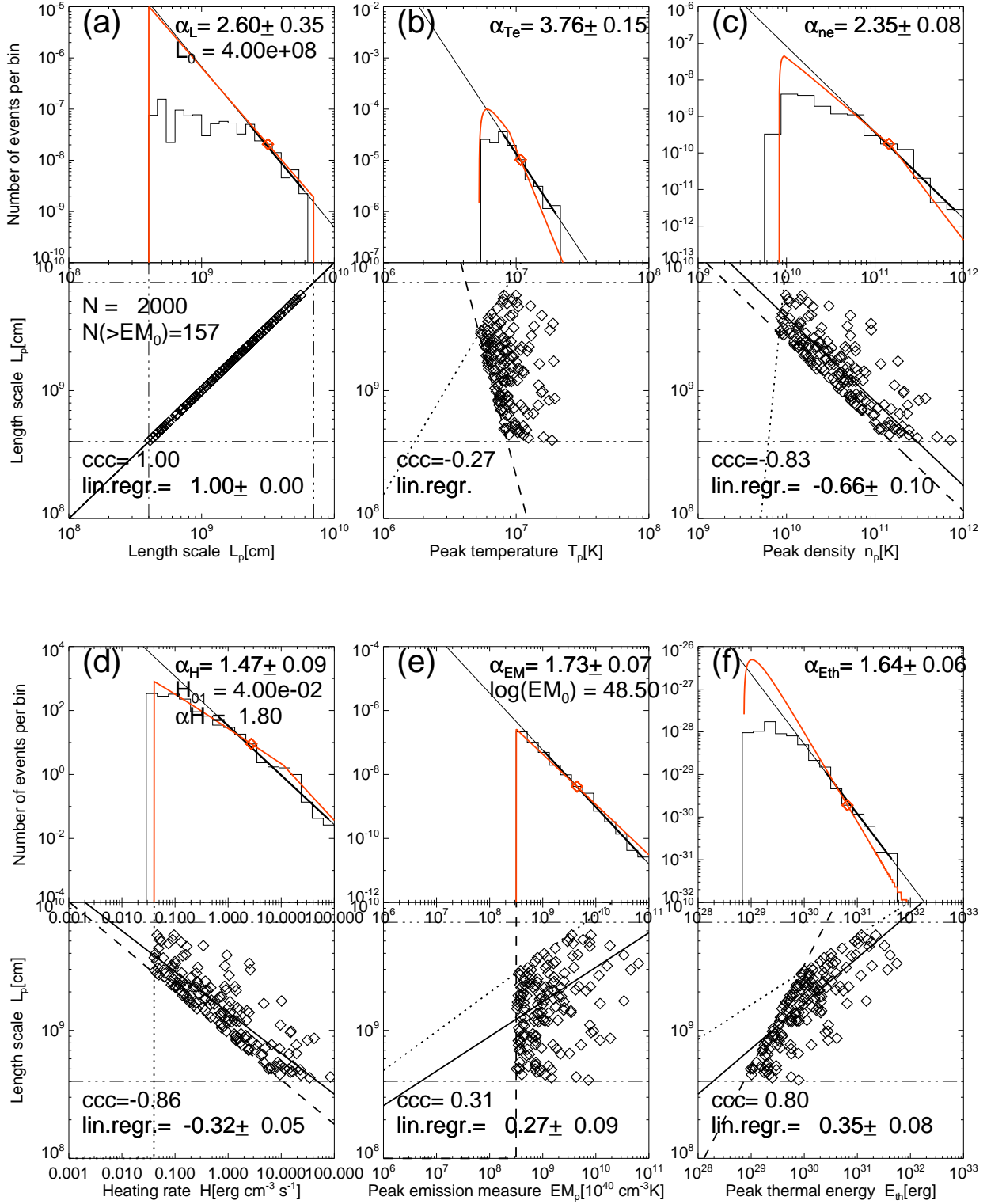}
\caption{Monte-Carlo simulations of data points (diamonds)
using the RTV relationships 
and a heating rate distribution $N(H) \propto H^{-\alpha_H}$ 
with a minimum value $H_0=0.4$ erg cm$^{-3}$ s$^{-1}$ and powerlaw slope
$\alpha_H=1.8$, an emission measure threshold of 
$EM_0 \ge 10^{48.5}$ cm$^{-3}$. The size distributions derived from 
analytical calculations (see Appendix A) are overlaid with red curves.
Otherwise similar representation as for the observed data in Fig.~7).}
\end{figure}

\begin{figure}
\plotone{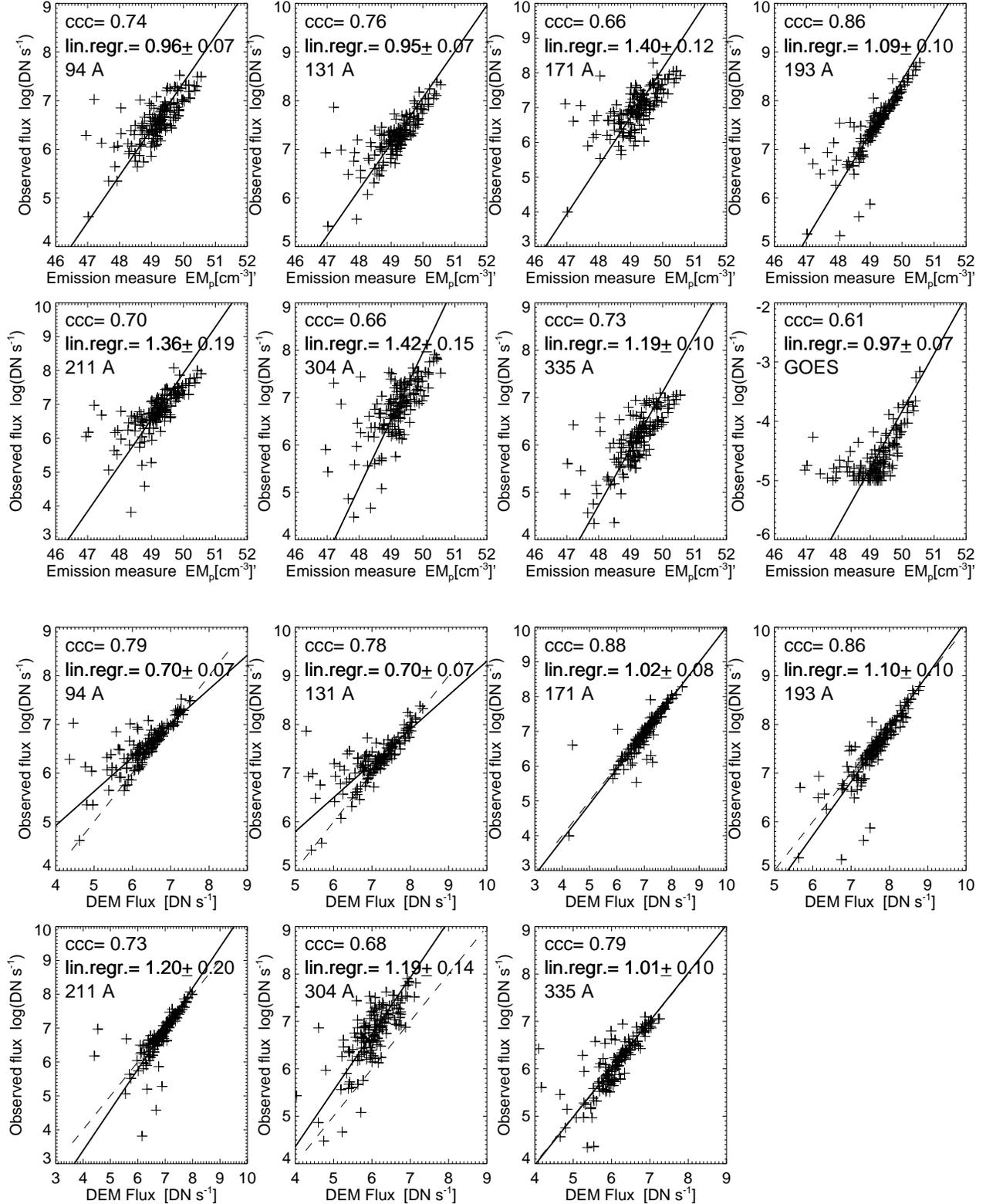}
\caption{{\sl Upper half:} Correlations between the observed fluxes 
$F_{\lambda}$ of the 7 AIA channels and the soft GOES channel with 
the fitted peak emission measure $EM_p$ peaks of the 155 analyzed
flares. {\sl Bottom:} Correlations between the observed fluxes
$F_{\lambda}$ and the predicted fluxes $F_{DEM}$ based on the
Gaussian DEM peak fits.}
\end{figure}

\begin{figure}
\plotone{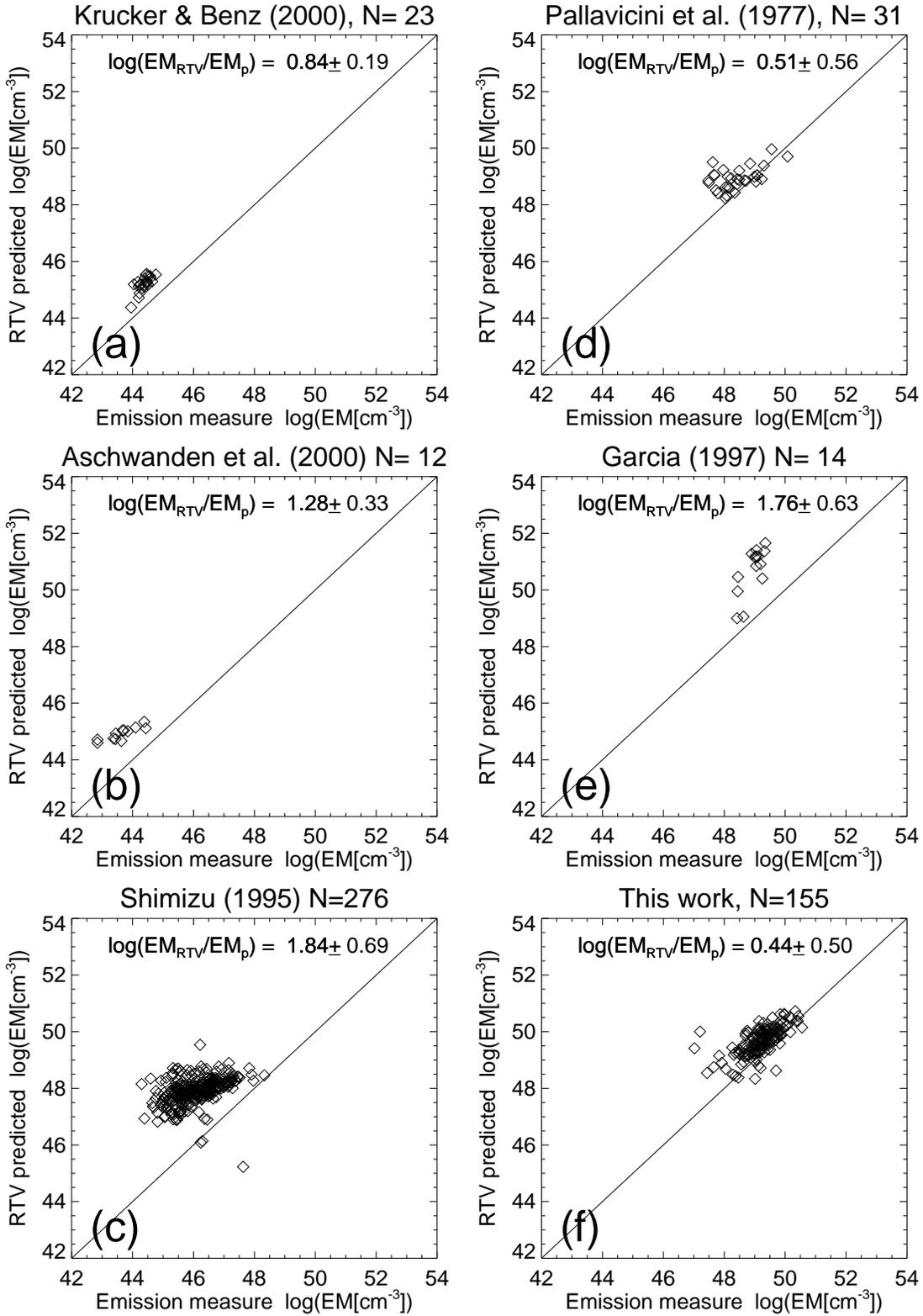}
\caption{Test of the Rosner-Tucker-Vaiana (RTV) scaling law with
previous measurements, including: (a) EUV nanoflares observed with
SOHO/EIT (Krucker and Benz 2000); (b) EUV nanoflares observed with
TRACE (Aschwanden et al.~2000); (c) active region transient
brightenings observed with Yohkoh/SXT (Shimizu 1995); (d) flares
observed with Skylab (Pallavicini et al.~1977); flares observed
with GOES and Yohkoh/SXT (Garcia 1997); and (e) flares observed
with AIA/SDO (this work).}
\end{figure}

\begin{figure}
\plotone{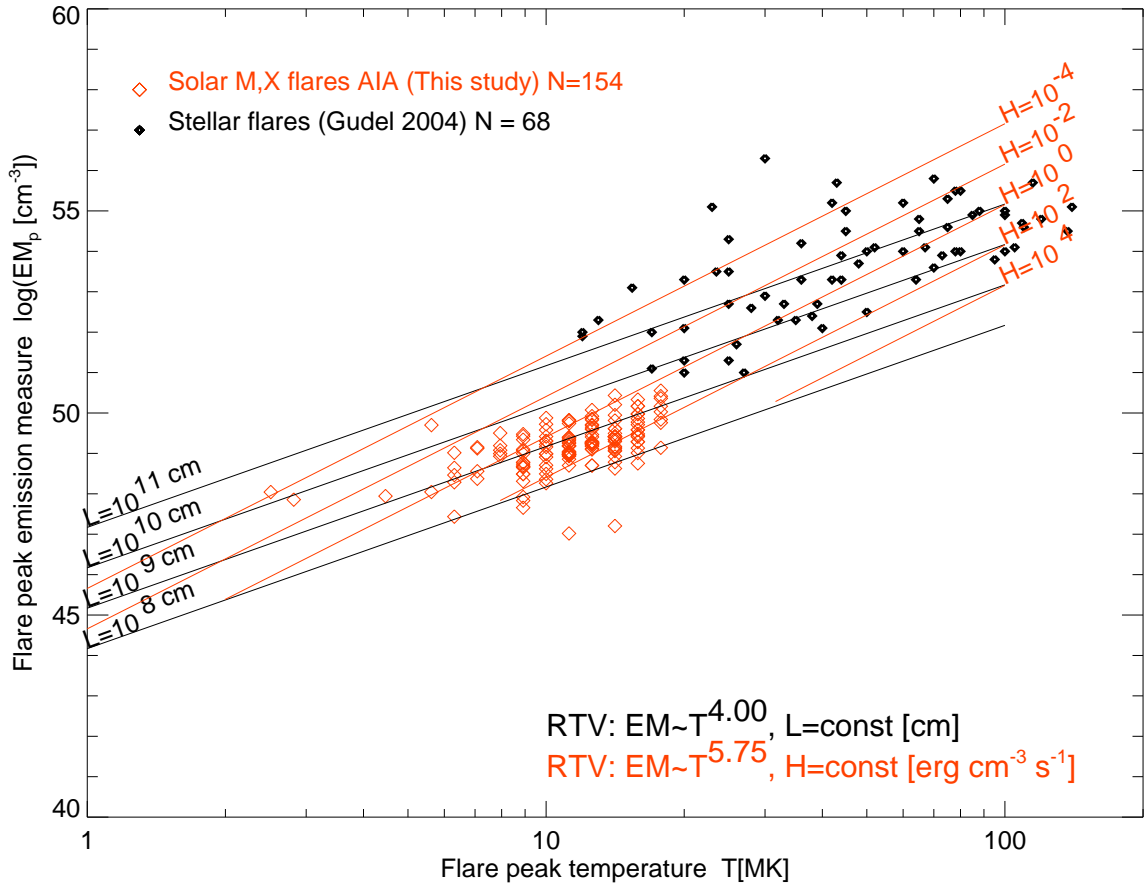}
\caption{Scatterplot of flare total emission measure $EM_p$ versus
the flare peak temperature $T_p$ for solar and stellar data sets.
The predictions of the RTV law are indicated for constant loop
lengths $L$ (black lines) and for constant heating rates $H$
(red lines).}
\end{figure}

\begin{figure}
\plotone{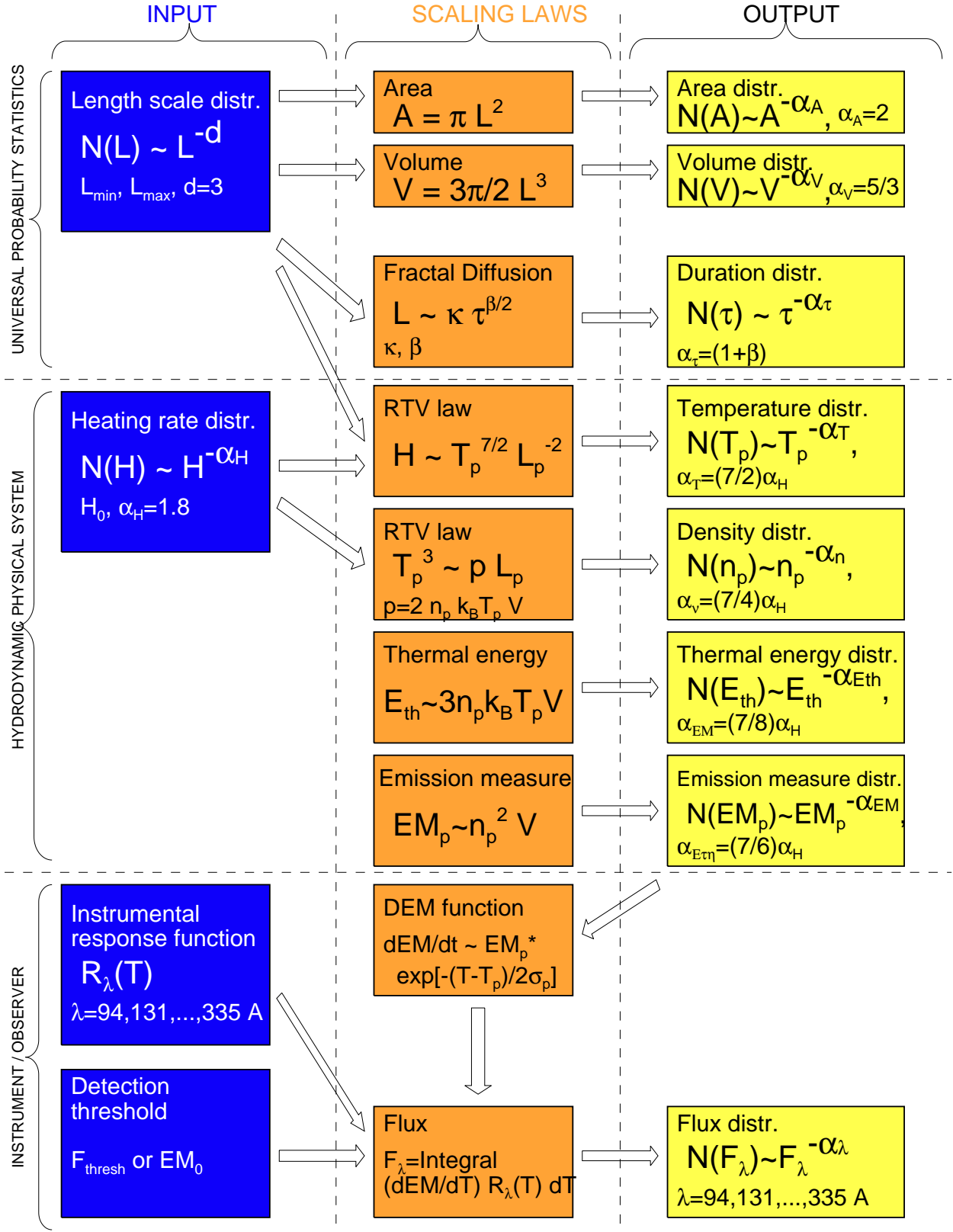}
\caption{Flow chart of input parameters (left), scaling laws (middle),
and output distribution functions (right) of the fractal-diffusive
SOC model applied to solar flares. The spatio-temporal parameters
($L, A, V, \tau$) follow from universal probability statistics
(top part of diagram), while the physical parameters and their
scaling laws are specific to the hydrodynamics of solar flares
(middle part of diagram), and the instrumental response functions
as a function of temperature and wavelengths are specific to the
observer (bottom part of diagram). The given powerlaw indices $\alpha_x$
are approximative values for dimensionality $d=3$.}
\end{figure}

\end{document}